\documentclass[twocolumn]{aastex631}

\usepackage{amsmath,amstext}
\usepackage[T1]{fontenc}
\usepackage{multirow}
\usepackage[utf8]{inputenc} 
\usepackage{float}
\usepackage{graphicx}

\usepackage{tabularx} 

\usepackage{threeparttable}

\shorttitle{Updated Mass, Eccentricity, and Tidal Heating of LP 791-18 d}
\shortauthors{Greklek-McKeon et al.}

\begin{document}

\title{Updated Mass, Eccentricity, and Tidal Heating Constraints for the Earth-sized Planet LP 791-18 d}

\correspondingauthor{Michael Greklek-McKeon}
\email{michael@caltech.edu}

\author[0000-0002-0371-1647]{Michael Greklek-McKeon}
\affiliation{Division of Geological and Planetary Sciences, California Institute of Technology, Pasadena, CA 91125, USA}

\author[0000-0002-5375-4725]{Heather A. Knutson}
\affiliation{Division of Geological and Planetary Sciences, California Institute of Technology, Pasadena, CA 91125, USA}

\author[0000-0002-1422-4430]{W. Garrett Levine}
\affiliation{Department of Astronomy, Yale University, New Haven, CT 06511, USA}

\author[0000-0003-2215-8485]{Renyu Hu}
\affiliation{Jet Propulsion Laboratory, California Institute of Technology, 4800 Oak Grove Drive, Pasadena, CA 91109, USA}
\affiliation{Division of Geological and Planetary Sciences, California Institute of Technology, Pasadena, CA 91125, USA}

\author[0000-0001-9518-9691]{Morgan Saidel}
\affiliation{Division of Geological and Planetary Sciences, California Institute of Technology, Pasadena, CA 91125, USA}

\author[0000-0002-0672-9658]{Jonathan Gomez Barrientos}
\affiliation{Division of Geological and Planetary Sciences, California Institute of Technology, Pasadena, CA 91125, USA}

\author[0000-0002-7094-7908]{Konstantin Batygin}
\affiliation{Division of Geological and Planetary Sciences, California Institute of Technology, Pasadena, CA 91125, USA}

\author[0000-0001-5578-1498]{Björn Benneke}
\affiliation{Department of Physics and Trottier Institute for Research on Exoplanets, Universit´e de Montr´eal, Montreal, QC, Canada}

\begin{abstract}

LP 791-18 d is a temperate Earth-sized planet orbiting a late M dwarf, surrounded by an interior super-Earth (LP 791-18 b, $R_P$ = 1.2 $R_{\oplus}$, $P=0.95$ days) and an exterior sub-Neptune (LP 791-18 c, $R_P$ = 2.5 $R_{\oplus}$, $P=4.99$ days). Dynamical interactions between LP 791-18 d and c produce transit timing variations (TTVs) that can be used to constrain the planet masses and eccentricities. These interactions can also force a non-zero eccentricity for LP 791-18 d, which raises its internal temperature through tidal heating and could drive volcanic outgassing. We present three new transit observations of LP 791-18 c with Palomar/WIRC, including the most precise TTV measurements ($<$ 6 seconds) of this planet to date. We fit these times with a TTV model to obtain updated constraints on the mass, eccentricity, and tidal heat flux of LP 791-18 d. We reduce the mass uncertainty by more than a factor of two ($M_d$ = 0.91 $\pm$ 0.19 $M_{\oplus}$). We perform an updated fit assuming tidally damped free eccentricities and find $e_d = 0.0011^{+0.0010}_{-0.0008}$ and $e_c = 0.0001 \pm 0.0001$, consistent with circular orbits. We find that the observed TTVs are not sensitive to $e \leq$ $\sim$0.01. Without a tidally damped eccentricity prior, $e_d = 0.056^{+0.015}_{-0.014}$, much higher than the eccentricity predicted by n-body simulations incorporating the effects of dynamical excitation and tidal damping. We predict the timing of upcoming JWST secondary eclipse observations for LP 791-18 d, which could tightly constrain the eccentricity and tidal quality factor of this Earth-sized exoplanet.

\end{abstract}

\section{Introduction} \label{sec:intro}

Small planets transiting low-mass stars enable the detailed study of exoplanets with Earth-like sizes and compositions. Of these planets, temperate worlds transiting cooler and smaller late M stars offer the rare opportunity to study the atmospheres of these Earth-sized planets with the James Webb Space Telescope (JWST). Planets orbiting smaller stars have relatively stronger atmospheric transmission features, making even compact, high mean-molecular weight atmospheres accessible to characterization with JWST \citep[e.g.][]{Piaulet2024,Banerjee2024,Gressier2024}. Yet, despite extensive ongoing searches, there are currently only four known temperate ($<$ 400 K), Earth-sized (< 1.1 $R_{\oplus}$) transiting planets outside of the TRAPPIST-1 system \citep{NASA_exoplanet_archive}. This population includes a recently discovered Earth-sized planet transiting the M6 star LP 791-18 \citep[][hereafter referred to as P23]{Peterson2023}. LP 791-18 d has a radius of 1.03 $\pm$ 0.04 $R_{\oplus}$, an orbital period of 2.75 days, and an equilibrium temperature of 396 K, with a tidally-locked permanent night-side that could plausibly allow for water condensation (P23). It is part of a three planet system, with a sub-Neptune (2.5 $R_{\oplus}$, P = 4.99 days) LP 791-18 c on an exterior orbit, and a super-Earth (1.2 $R_{\oplus}$, P = 0.95 days) LP 791-18 b on an interior orbit \citep{Crossfield_2019}.

It is often challenging to measure masses for small planets using the radial velocity technique \citep[e.g.][]{Dai2024,Wright2018,Fischer2016}. Fortunately, dynamical interactions between planets in the LP 791-18 system cause detectable transit timing variations (TTVs) that can be used to constrain the planet masses and orbital eccentricities. P23 carried out an extensive transit follow-up campaign to characterize the TTVs in this system, including a near-continuous 172 hour Spitzer observation that was conducted in 2019 to confirm the existence of LP 791-18 d. These data were supplemented by an additional 19 ground-based transits of planet c and 40 transits of planet d executed on small and mid-sized telescopes including LCOGT, EDEN, ExTrA, SPECULOOS, TRAPPIST, MEarth, and MuSCAT. These observations resulted in the detection of TTV signals with amplitudes of $\sim$2.5 min and $\sim$0.5 min for for planets c and d, respectively. P23 also found that there was a strong “chopping” TTV signal in their data, indicating that the planets in this system have relatively large mass ratios and/or eccentric orbits \citep{Lithwick_2012,TTVFast}. They found that LP 791-18 c has a mass of 7.1 $\pm$ 0.7 $M_{\oplus}$ and LP 791-18 d has a mass of 0.9$^{+0.5}_{-0.4}$ $M_{\oplus}$, and predicted with dynamical simulations that LP 791-18 d maintains a non-zero forced eccentricity of $\sim$0.0015 that is stable long-term over the system age due to dynamical interactions with the nearby orbit of the more massive LP 791-18 c. 

LP 791-18 d's relatively close-in orbit and large planet-star mass ratio means that this tiny eccentricity of 0.0015 has an outsized impact on the planet's overall energy budget. Tidal heating in rocky planets impacts interior ice and silicate melting, surface temperature, atmospheric properties, and potential habitability \citep{Jackson2008, Quick2020, Seligman2023}. If LP 791-18 d has an Earth-like composition and tidal quality factor, the implied tidal heat flux from an eccentricity of 0.0015 is similar to that of Jupiter's moon Io, the most volcanically active body in the solar system (P23). This has significant implications for the atmosphere of LP 791-18 d, which may be dominated by volcanically produced gases. Recent JWST/NIRSpec/NIRISS transmission spectroscopy of the 1.58 $R_{\oplus}$ planet L 98-59 d indicated that this planet may host a sulfur-rich atmosphere, potentially due to volcanic outgassing \citep{Gressier2024,Banerjee2024}. Like LP 791-18 d, L 98-59 d also has a non-zero eccentricity (0.074$^{+0.057}_{-0.046}$) that has been suggested as a potential driver of volcanic outgassing due to tidal heating \citep{Seligman2023}. An upcoming JWST/MIRI program (GO 6457) will search for the presence of an outgassed atmosphere for LP 791-18 d using secondary eclipse observations, but the predicted tidal heat flux and volcanic outgassing rate depends in part on the orbital eccentricity, and the predicted timing of these eclipses depends sensitively on the assumed orbital eccentricity and longitude of periastron.

The precision of the current mass and eccentricity constraints for LP 791-18 d are limited by the relatively small amplitude of the TTVs ($\sim$ 40~s) that it induces on the much more massive planet c. Among the ensemble of transit timing measurements presented in P23, the Spitzer observations were the only ones precise enough to detect these timing variations, and as a result they drive the TTV constraints for both planets. In this study, we utilize diffuser-assisted infrared transit photometry with the Wide-field InfaRed Camera \citep[WIRC,][]{Vissapragada2020} at Palomar Observatory to observe three new transits of LP 791-18 c.  With this setup, we achieved a timing precision superior to that of Spitzer.  In \S\ref{sec:Transit follow-up}, we describe our transit follow-up observations. In \S\ref{sec:ttv modeling}, we describe our TTV modeling and analysis. 
In \S\ref{sec:Discussion}, we discuss our results, and in \S\ref{sec:Conclusions} we summarize our key findings.

\section{Transit Follow-up} \label{sec:Transit follow-up}

\subsection{Palomar/WIRC Observations} \label{sec:Palomar observations}

We observed three transits of LP 791-18 c in the $J$-band with WIRC on the 200" Hale Telescope at Palomar Observatory, California, USA. The Hale Telescope is a 5.08-m telescope equipped with a 2048 x 2048 Rockwell Hawaii-II NIR detector, providing a field of view of 8\farcm7 × 8\farcm7 with a plate scale of 0.''25 per pixel \citep[WIRC,][]{Wilson2003}. Our data were taken with a beam-shaping diffuser that increased our observing efficiency and improved the photometric precision and guiding stability \citep{Stefansson2017,Vissapragada2020}. 

\begin{figure*}
\begin{center}
  \includegraphics[width=18cm]{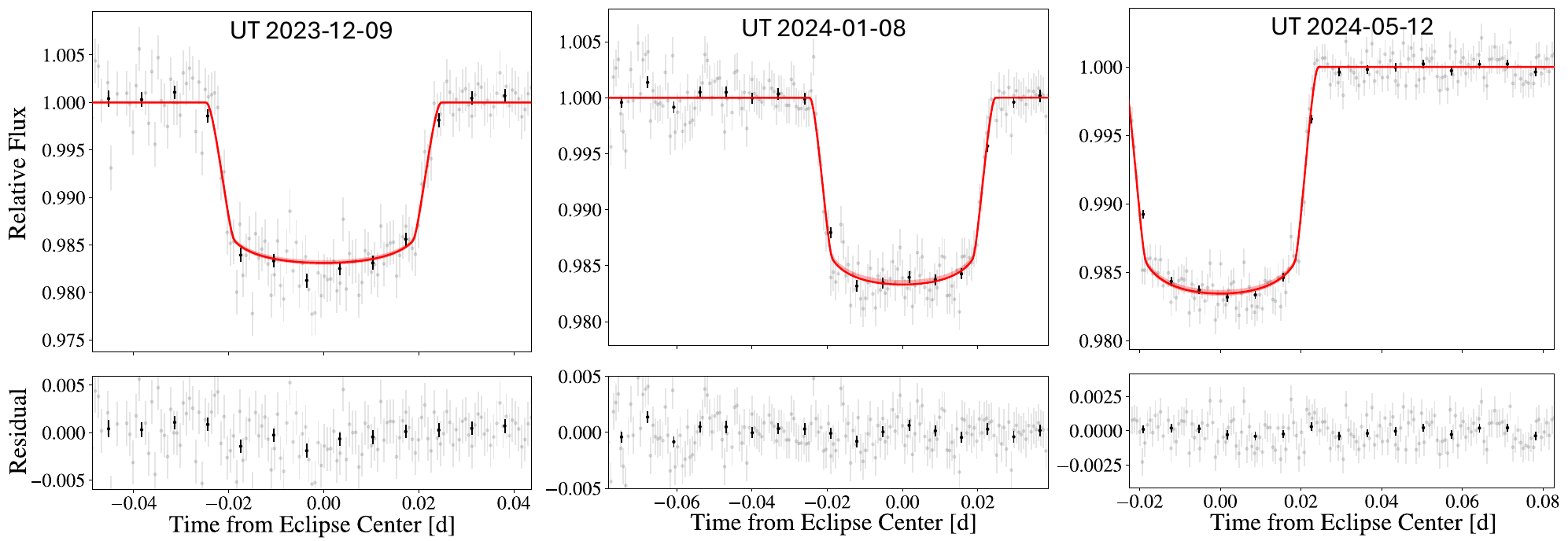}
  \caption{Detrended Palomar/WIRC light curves for the three transit observations of LP 791-18 c (upper panels) and residuals after the best-fit transit light curve has been subtracted (lower panels). Unbinned data are shown as grey circles, with 10 minute binned points overplotted as black circles.  The best joint-fit transit models are overplotted as red lines, with red shading to indicate the 1$\sigma$ uncertainties on the transit shape.}
  \label{fig:Palomar}
\end{center}
\end{figure*}

We observed transits of LP 791-18 c on UT 2023-12-09, 2024-01-08, and 2024-05-12. We used 10-second exposure times stacked to 4 total co-added exposures per image, and observed full transits plus at least 1 transit duration of baseline (Figure \ref{fig:Palomar}). Exposure time is typically varied on the night of observation based on the sky conditions (Moon fraction and proximity, cloud cover, etc.), but we found 40 second total exposure time to be sufficient for all observations. For each night, we obtained calibration images to dark-subtract, flat-field, remove dead and hot pixels, and remove detector structure with a 9-point dithered sky background frame following the methodology of \cite{Vissapragada2020}. We extracted photometry for our target star and a set of nearby comparison stars, and we chose up to 10 comparison stars that have minimal variance relative to the time-changing flux of the target star. For the UT 2024-01-08 and 2024-05-12 observations, these were the same 8 comparison stars, while for the UT 2023-12-09 observation we used 2 additional comparison stars for a total of 10. The final number of comparison stars was decided by the significance of their weights in the fitting procedure described in \S\ref{sec:transit analysis}. We cleaned the target and comparison light curves by applying a moving median filter with a width of 31 data points and removing 5$\sigma$ outliers. We then tested various target aperture sizes from 5 - 25 pixels and selected the optimal aperture by minimizing the root mean square scatter after the light-curve fitting described in \S\ref{sec:transit analysis}. Our optimal apertures were 19, 19, and 12 pixels for UT 2023-12-09, 2024-01-08 and 2024-05-12, respectively. For the UT 2023-12-09 observation we experienced poor seeing approaching the 3'' width of our diffuser and intermittent partial cloud cover, and for the UT 2024-01-08 observation we experienced poor seeing below 3''. This affected our photometric precision and target PSF width, but we do not exclude any data from these observations because the comparison star fluxes still track very closely with the target star flux resulting in strong transit detections (Figure \ref{fig:Palomar}). Additional information about our transit observations, including observation times, transit coverage, and airmasses, are provided in Table \ref{tab:Palomar obs}.

\begin{table*}
\centering
\caption{Summary of ground-based Palomar/WIRC observations of LP 791-18 c.}
\label{tab:Palomar obs}
\hspace*{-2cm}
\begin{tabular}{ccccccccc}
\hline
\hline
UT Date & Start & Finish & Transit \% & Baseline \% & z$_{\text{st}}$ &  z$_{\text{min}}$ & z$_{\text{end}}$ & Transit Midtime (BJD, \S\ref{sec:transit analysis})
\\ \hline
2023 Dec 9 & 10:31:42 & 12:44:43 & 100\% & 105\% & 2.54 & 1.60 & 1.60 & 2460287.987952 $\pm$ 0.000070
\\ 
2024 Jan 8 & 08:18:33 & 11:06:46 & 100\% & 140\% & 2.84 & 1.57 & 1.57 & 2460317.927413 $\pm$ 0.000050
\\ 
2024 May 12 & 03:32:42 & 06:06:23 & 97\% & 120\% & 1.55 & 1.55 & 2.15 &  2460442.675472 $\pm$ 0.000037
\\ 
\hline
\end{tabular}
\begin{tablenotes}
      \item[a] Start and Finish columns represent the time of first and last science images in UT time, the transit and baseline fractions are relative to the total transit duration for LP 791-18 c, z$_{\text{min}}$ is the minimum airmass of the science sequence while z$_{\text{st}}$ and z$_{\text{end}}$ are the starting and ending airmasses.
\end{tablenotes}
\end{table*}

\subsection{Transit Modeling} \label{sec:transit analysis}

We fit the WIRC light curves using \texttt{exoplanet} with a combined systematics and transit model as described in \cite{GreklekMcKeon2023}. Our systematics model for each night included a linear combination of comparison star light curve weights, an error inflation term added in quadrature to the flux errors, and a linear slope. We also tested systematics models with linear combinations of weights for the target centroid offset, PSF width, airmass, and local sky background as a function of time. We compared the Bayesian Information Criterion \citep[BIC,][]{Schwarz1978} for all possible combinations of these systematic noise parameters using the same framework as in \cite{Jorge2024}. We found that the model that produced the lowest BIC value included weights for the target PSF width on UT 2024-01-08, while our UT 2023-12-09 and 2024-05-12 observations preferred no additional detrending parameters in the systematics model. When optimizing the systematics model for each night, we also fit for the transit shape parameters (impact parameter $b$, planet-star radius ratio $R_P$/$R_*$, and semi-major axis ratio $a$/$R_*$), and found that they were all consistent within 1$\sigma$ across all three nights.

We fit the three WIRC transits jointly (Figure \ref{fig:Palomar}). We used the same model framework as in \cite{GreklekMcKeon2023}, with a wide uniform prior of $\pm$2 hours on the transit times, transit shape parameters (impact parameter $b$, planet-star radius ratio $R_P$/$R_*$, semi-major axis ratio $a$/$R_*$) shared across nights, with different systematics model parameters and comparison star weights for each night. We adopted stellar parameters from P23 and used \texttt{ldtk} to calculate the $J$-band quadratic WIRC limb darkening parameters $u_1=0.182$ and $u_2=0.151$, and held them fixed in our fits. We explored the parameter space with the NUTS sampler in \texttt{PyMC3} for 2500 tune and 2000 draw steps, and confirmed that the chains have evolved until the Gelman-Rubin statistic values are < 1.001 for all parameters. Our measured transit times are listed in Table \ref{tab:Palomar obs}, and the final transit light curves are shown in Figure \ref{fig:Palomar}. We achieved transit timing precisions of 6, 4, and 3 seconds for our Palomar/WIRC transits of LP 791-18 c.  This represents an improvement over the 9 second transit timing precision  of the two Spitzer transits of planet c presented in P23. 

\section{TTV Modeling}  \label{sec:ttv modeling}

\subsection{Validation of TTV Model and Choice of Eccentricity Parameterization} \label{sec:TTV parameterization}

Before performing an updated TTV model fit using our new transit times, we first fit the original set of transit times listed in Extended Data Tables 1 and 2 of P23, where we independently reproduced their solution. We used the \texttt{TTVFast} package to model the observed transit times. \texttt{TTVFast} \citep{TTVFast} is a computationally efficient $n$-body code that uses a symplectic integrator with a Keplerian interpolator to calculate transit times in multi-planet systems. The modeled transit times are a function of the planetary mass ratios and orbital elements relative to a reference epoch, which we chose to be $T_0 = 8546.0$ (BJD - 2450000), the same as P23. In our TTV modeling, we also fixed the planetary orbital inclinations ($i$) to 90$^{\circ}$, since the transit fits indicate that planets d and c have a low mutual inclination and are very close to edge-on ($i_d$ = 89.34 $\pm$ 0.41, $i_c$ = 89.78 $\pm$ 0.13, P23). 
For a purely edge-on orbital inclination the longitude of the ascending node ($\Omega$) becomes undefined, so we arbitrarily set it to 90$^{\circ}$ for both planets. The TTV model also includes the planet-to-star mass ratios ($M_p$/$M_{*}$), Keplerian orbital periods ($P$), mean anomalies reparamaterized with the time of first transit ($t_0$), and the planetary eccentricities ($e$) and longitudes of periastron ($\omega$). This results in a total of ten free parameters in our 2-planet TTV model. Prior distributions used for these model parameters are described in Table \ref{tab:results}. Also as in P23, we did not include planet b in the TTV analysis because the predicted amplitude of its TTVs and its predicted impact on the TTVs of planets c and d are less than one second, independent of the planet masses. Following P23, we initially fit the data with $e$ and $\omega$ parameterized as $e$cos($\omega$) and $e$sin($\omega$) \citep{exofast}. As originally noted by \cite{Ford2006}, fitting for $e$sin($\omega$) and $e$sin($\omega$) results in an effective linear prior on $e$, which must be corrected by weighting the stepping probability by $e_{i-1}$/$e_i$, where the subscript $i$ denotes the current link in the chain. This approach has been used in previous TTV studies \citep[e.g.][]{Agol_2021}, and was the approach adopted by P23, so we use it as well.

The orbital periods, mean anomalies, eccentricities, and longitudes of periastron are osculating orbital elements defined at the TTV model start time $T_0$. Along with this TTV model parameterization, we incorporated the damped-state free eccentricity prior described in P23. This study performed long-term n-body simulations with \texttt{REBOUND} \citep{REBOUND}, which indicated that the eccentricities of the planets should be damped to values near zero ($\sim$0.001 for d, and $\sim$0.0001 for c) on relatively short timescales (see \S\ref{sec:Damped TTVs} for additional analysis of this). These damped-state eccentricities oscillate around stable long-term equilibria, and represent the forced eccentricities that are preserved due to mutual gravitational interactions between the planets after the free eccentricity has been stripped away by tidal damping. P23 implemented this damped-state prior on the eccentricities by computing the free and forced eccentricity components for planets c and d for each proposed step in the MCMC chain using a two year \texttt{REBOUND} simulation, and then imposing Gaussian priors on the values of the free eccentricities centered at zero and with a standard deviation of 0.001 for planet d and 0.0001 for c. We found that we obtained equivalent results using \texttt{REBOUND} simulations with a duration of 6 months rather than the two years utilized by P23 and adopted this approach in order to decrease the run time of our fits. This had the added advantage of making sure that our estimate of the forced eccentricity was not biased by the long-term osculations of the free eccentricity term, while still accurately separating the free versus forced eccentricity components by simulating many of the 26-day TTV super-periods of LP 791-18 d and c (P23).

We fit our TTV model to the data from P23 using the affine invariant Markov chain Monte Carlo (MCMC) ensemble sampler \texttt{emcee} \citep{emcee}, and chose wide uniform priors for all parameters: $\textit{U}$(-1, 1) for $ecos(\omega)$ and $esin(\omega)$, $\textit{U}$(0, 30$M_{\oplus}$) for the planet-star mass ratios, $\textit{U}$(-100$\sigma$, +100$\sigma$) for the planetary orbital periods and $t_{t_0}$ values using the $\sigma$ values reported in P23, in addition to the damped-state free eccentricity prior. We initialized the MCMC fit with 200 walkers (20 per free parameter) randomly distributed across the full prior volume. We ran the sampler for 7000 steps, which is more than 50 autocorrelation lengths for all parameters after discarding the initial 1500 steps as burn-in. We obtained results consistent within 1$\sigma$ to those reported in P23 for all fit parameters with this damped-state eccentricity model framework. When we repeated the TTV analysis without the damped-state eccentricity prior, we also obtained planetary mass values consistent with those reported in P23.

\begin{figure}
  \includegraphics[width=8.5cm]{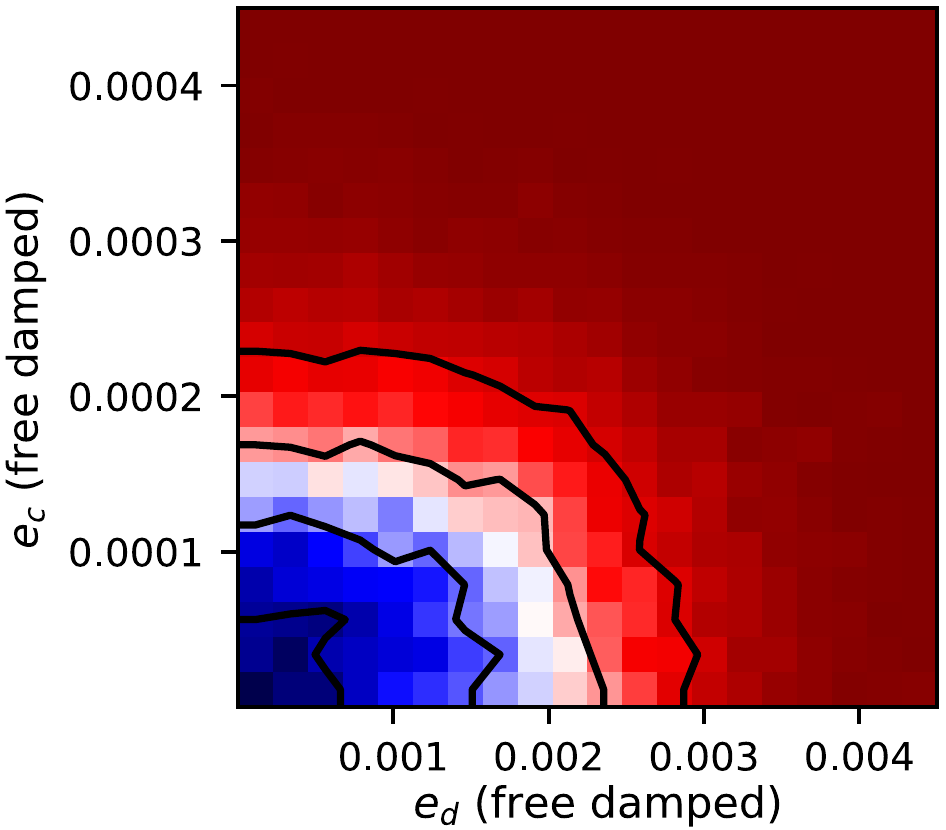}
  \caption{The posterior probability density distribution of planetary eccentricities for LP 791-18 d and c with a fit to the TTV data of P23 using a tidally damped-state eccentricity prior and parameterizing $e$ and $\omega$ as $\sqrt{e}\cos(\omega)$ and $\sqrt{e}\sin(\omega)$.}
  \label{fig:P23_Ecc_dist}
\end{figure}

Next, we repeated our fit to the TTV data from P23 with an alternative parameterization for $e$ and $\omega$.  
 correction method used to eliminate the effective positive linear prior in $e$ for the $e$cos($\omega$) and $e$sin($\omega$) parameterization described in \cite{Ford2006} and utilized by P23 preferentially rejects steps with higher eccentricity and results in an approximately but not perfectly uniform prior in $e$. Due to the singularity at $e = 0$, there is a very slight overcorrection as $e$ approaches zero, as can be seen in \cite{exofast} Figure 1. For this reason, \cite{exofast} recommend using the $\sqrt{e}\cos(\omega)$ and $\sqrt{e}\sin(\omega)$ parameterization, which naturally recovers a uniform prior in $e$ and $\omega$. When we repeated our fit to the P23 data using this parameterization, we found that the small but non-zero eccentricities reported by P23 ($e_d$ = 0.0015 $\pm$ 0.00014 and $e_c$ = 0.0008 $\pm$ 0.0004) for planets d and c vanished and we instead retrieved eccentricity posteriors consistent with zero for both planets (Figure \ref{fig:P23_Ecc_dist}). When we compared this tidally damped eccentricity fit to one where we forced $e$ = 0 for both planets, the difference in best-fit predicted transit times is less than 1 second for both planets, compared to our smallest transit timing precision of 3 seconds. This confirms that the observational data are not sensitive to eccentricities this small ($\sim$0.001), and the originally reported values from P23 were an artifact of the chosen $e$ and $\omega$ parameterization. In order to avoid this bias, we use the $\sqrt{e}\cos(\omega)$ and $\sqrt{e}\sin(\omega)$ parameterization in all of our TTV model fits going forward. 

\begin{figure*}
\begin{center}
  \includegraphics[width=18cm]{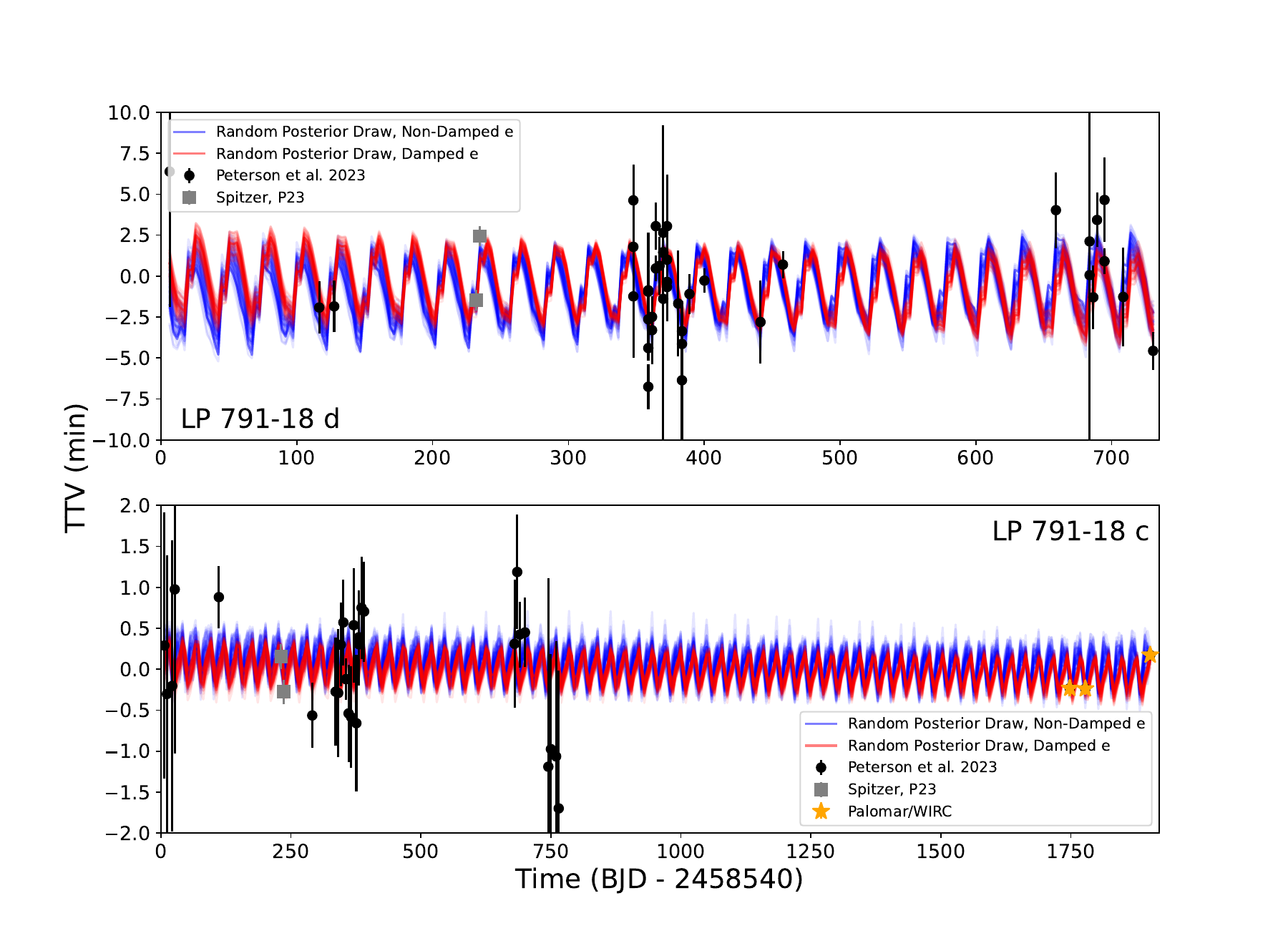}
  \caption{TTVs from the ground-based follow-up campaign of P23 (black circles) including the high-precision Spitzer observations (gray squares), and our new Palomar/WIRC follow-up observations (orange stars), with 100 random posterior draws from our damped eccentricity TTV model (red) and free eccentricity TTV model (blue) for LP 791-18 c (top panel) and d (bottom panel). Our Palomar/WIRC timing measurements for LP 791-18 c are the most precise TTV observations of this system so far, a factor of 1.5-3x more precise than Spitzer, and drive the TTV model constraints for LP 791-18 d.}
  \label{fig:TTVFig}
\end{center}
\end{figure*}

\subsection{The Damped Case} \label{sec:Damped TTVs}

We performed an updated TTV fit using the TTV observations from P23 and our three new transit observations of LP 791-18 c, applying the same free eccentricity tidally damped-state prior described in \S\ref{sec:TTV parameterization}. We ran the TTV MCMC retrieval with 200 walkers for 25,000 steps, ensuring that all parameters achieved $>$ 50 autocorrelation lengths. We show the updated suite of TTV observations in Figure \ref{fig:TTVFig}, with a representative sample of TTV model fits overplotted and a comparison to the non-damped free eccentricity TTV retrieval shown.  We found that the TTV-based eccentricities remain consistent with zero when applying the damped-state eccentricity prior (see Figure \ref{fig:MassEccFig}). This has significant implications for the tidal heat flux of LP 791-18 d, as discussed in \S\ref{sec:tidal heating}. Our updated planet masses are consistent with those reported in P23, but with a significantly improved mass uncertainty for LP 791-18 d.  We discuss the corresponding implications for the interpretation of this planet's bulk density in \S\ref{sec:bulk densities}.

\subsection{The Non-Damped Case} \label{sec:non-damped}

In order for the tidal circularization timescale \cite[][equation 1]{Puranam2018} to be comparable to or longer than the age of the system (> 0.5 Gyr, P23), the reduced tidal quality factor $Q'$ (defined as $Q$/$k_2$) must be $\geq 10^4$ for LP 791-18 d. This value would be similar to the super-Earth GJ 876 d, whose non-zero eccentricity is maintained by chaotic excitations from nearby planets \citep{Puranam2018}, but much larger (and less dissipative) than the values commonly assumed for terrestrial planets \citep[10-100,][]{Clausen2015}. For LP 791-18 c, $Q_c \geq 7\times10^3$ would yield a circularization time comparable to the system age. This is consistent with predictions of the range of $Q_p$ values for sub-Neptune exoplanets from interior modeling and population-level analysis \citep[e.g.][]{Goldreich1996,Nettelmann2011,Clausen2015,Tobie2019,Louden2023} and smaller (more dissipative) than measured values for Neptune-sized exoplanets \citep[e.g. GJ 436 b, $Q_p$ = $2 \times 10^5 - 10^6$,][]{Morley2017}. We caution that if LP 791-18 is considerably older than 1 Gyr, both planets would require higher $Q_p$ values and much less efficient tidal dissipation to retain any free eccentricity.  It is therefore likely that LP 791-18 d has a damped free eccentricity based on the predictions of $Q_p$ values for terrestrial planets, but a much larger $Q_p$ also cannot be ruled out. 

\begin{figure*}
\begin{center}
  \includegraphics[width=18cm]{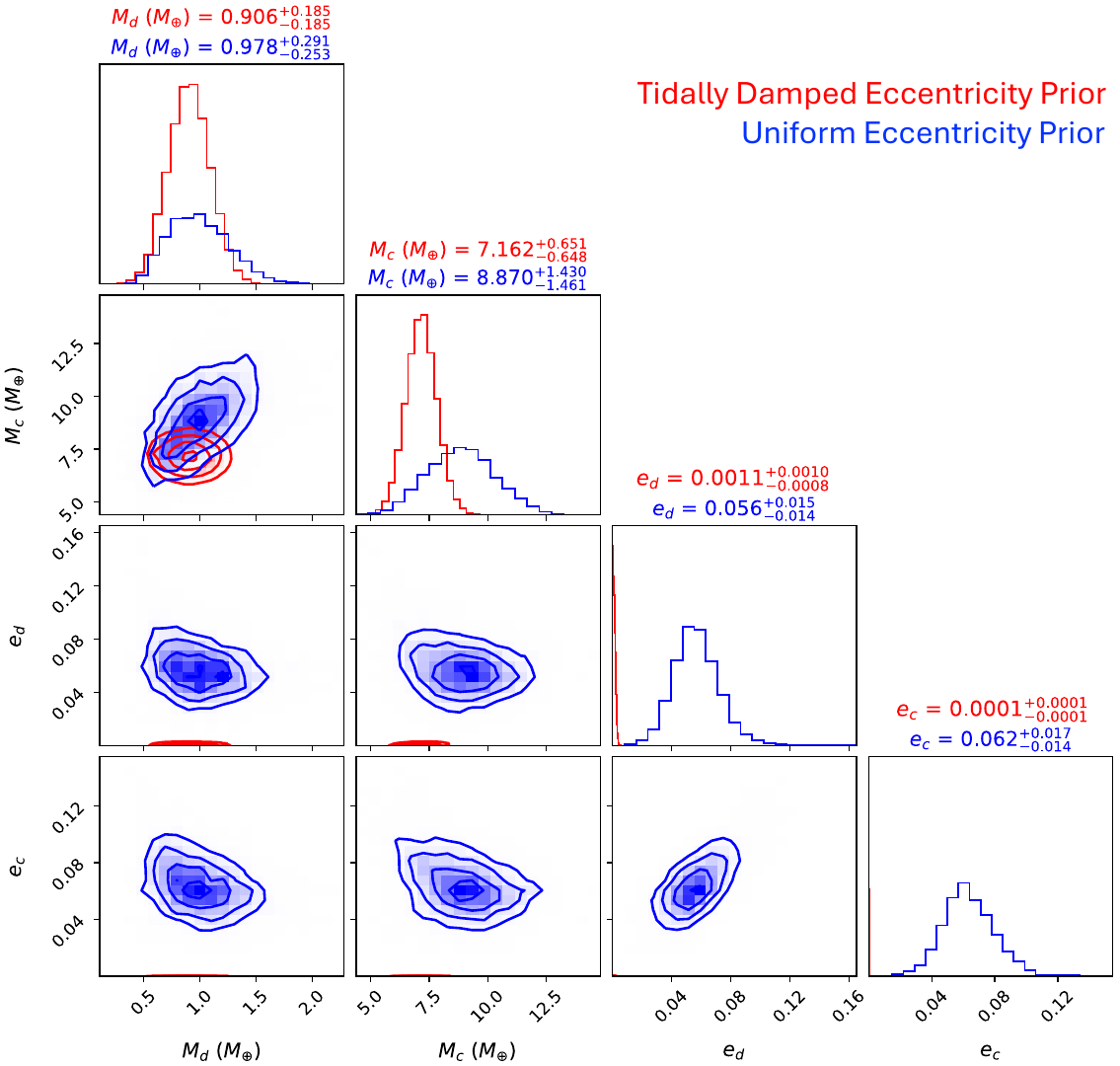}
  \caption{Posterior distributions for the masses and eccentricities of LP 791-18 d and c from a fit to the updated set of TTV observations in the case of a tidally damped free eccentricity prior (red), and a uniform eccentricity prior (blue). If we assume that the eccentricities are damped, then the observations are consistent with $e = 0$ and the planetary masses are slightly lower. If the eccentricities are not damped, the planetary masses are slightly higher.}
  \label{fig:MassEccFig}
\end{center}
\end{figure*}

We explore this possibility by performing an updated TTV fit including our Palomar/WIRC observations for the case with a free eccentricity prior.  In this case, we used the same TTV model parameterization as before, with a uniform prior from -1 to 1 on $\sqrt{e}\cos(\omega)$ and $\sqrt{e}\sin(\omega)$, but without the additional damped-state free eccentricity prior. We ran this MCMC retrieval with 200 walkers for 5$ \times 10^5$ steps, ensuring that our chains extended over more than 50 autocorrelation lengths for each parameter. The results of this TTV model and the damped-state eccentricity prior TTV retrieval are shown in Figure \ref{fig:MassEccFig}.

We find that our undamped fit prefers moderately non-zero eccentricity values for both planets ($e_d$ = 0.056$^{+0.015}_{-0.014}$ and $e_c$ = 0.062$^{+0.017}_{-0.014}$). These higher eccentricities also result in slightly higher masses for both planets, although with larger mass uncertainties and still consistent within $1\sigma$ to the fit with a damped eccentricity prior (Figure \ref{fig:MassEccFig}). This slight shift in planet masses when using a less restrictive eccentricity prior is consistent with the known degeneracy between planet mass and eccentricity in TTV fits \citep{Lithwick_2012}. 

The current data cannot differentiate between the damped versus undamped models. We compare the Bayesian Information Criterion \citep[BIC,][]{Schwarz1978} for the maximum likelihood model parameters in the damped versus non-damped eccentricity TTV models, and found that the non-damped TTV model is preferred with a $\Delta$BIC of -1.3, an insignificant difference using the $\Delta$BIC interpretation recommendations of \cite{raftery1995}. We discuss the implications of these potentially higher eccentricity and mass values for the tidal heat fluxes and compositions of the planets in \S\ref{sec:Discussion}.

\begin{table*}
  \centering
  \caption{Priors and posteriors for LP 791-18 model parameters.\label{tab:results}}
  \begin{tabular}{lccc}
    \hline
    \hline
    Parameter & Prior & \multicolumn{2}{c}{Posterior} \\
    \hline
    \multicolumn{4}{c}{\emph{Measured planetary parameters, Palomar/WIRC transit fit (\S\ref{sec:transit analysis})}} \\
    \hline
    && \multicolumn{2}{c}{\emph{LP 791-18 c}} \\
    $a/R_\star$ & $p(a/R_\star\vert P,M_{\star}, R_{\star})$ & \multicolumn{2}{c}{$37.05^{+0.05}_{-0.07}$} \\
    $R_p/R_\star$ & $\mathcal{U}(0.0,0.2)$ & \multicolumn{2}{c}{$0.12518^{+0.00027}_{-0.00027}$} \\
    Impact parameter, $b$ & $\mathcal{U}(0,1+R_p/R_\star)$ & \multicolumn{2}{c}{$0.0367^{+0.0323}_{-0.0222}$} \\
    \hline
    \multicolumn{4}{c}{\emph{Measured planetary parameters, TTV fit, Non-damped free eccentricity prior (\S\ref{sec:non-damped})}} \\
    \hline
    && \emph{LP 791-18 d} & \emph{LP 791-18 c} \\
    $P$ (days) & $\mathcal{U}($P$ - 1.0,$P$ + 1.0)$$^\textsc{1}$ & $2.75519^{+0.00044}_{-0.00046}$ & $4.98987^{+0.00006}_{-0.00007}$ \\
    $t_0$ (BJD - 2,458,546) & $\mathcal{U}(t_0 - 2.5,t_0 + 2.5)$$^\textsc{2}$ & $6.37773^{+0.00068}_{-0.00065}$ & $6.50935^{+0.00011}_{-0.00011}$ \\
    $\sqrt{e}\cos{\omega}$ & $\mathcal{U}(-1,1)$ & $-0.120^{+0.056}_{-0.048}$ & $-0.239^{+0.035}_{-0.035}$ \\
    $\sqrt{e}\sin{\omega}$ & $\mathcal{U}(-1,1)$ & $0.115^{+0.056}_{-0.086}$ & $0.039^{+0.059}_{-0.076}$ \\
    $M_p/M_\star$ x $10^{-6}$ & $\mathcal{U}(0,90)$ & $2.936^{+0.087}_{-0.076}$ & $26.63^{+4.29}_{-4.39}$ \\
    \hline
    \multicolumn{4}{c}{\emph{Measured planetary parameters, TTV fit, Damped-state free eccentricity prior (\S\ref{sec:Damped TTVs}) }} \\
    \hline
    && \emph{LP 791-18 d} & \emph{LP 791-18 c} \\
    $P$ (days) & $\mathcal{U}($P$ - 1.0,$P$ + 1.0)$$^\textsc{1}$ & $2.75485^{+0.00013}_{-0.00013}$ & $4.9899100^{+0.0000012}_{-0.0000014}$ \\
    $t_0$ (BJD - 2,458,546) & $\mathcal{U}(t_0 - 2.5,t_0 + 2.5)$$^\textsc{2}$ & $6.37888^{+0.00037}_{-0.00038}$ & $6.509230^{+0.000065}_{-0.000064}$ \\
    $\sqrt{e}\cos{\omega}$ & $\mathcal{U}(-1,1)$ & $-0.0033^{+0.0266}_{-0.0262}$ & $-0.0001^{+0.0008}_{-0.0008}$ \\
    $\sqrt{e}\sin{\omega}$ & $\mathcal{U}(-1,1)$ & $0.0034^{+0.0268}_{-0.0274}$ & $-0.00003^{+0.00754}_{-0.00769}$ \\
    $M_p/M_\star$ x $10^{-6}$ & $\mathcal{U}(0,90)$ & $2.719^{+0.056}_{-0.056}$ & $21.50^{+1.95}_{-1.95}$ \\    
    \hline
    \multicolumn{4}{c}{\emph{Derived planetary parameters }} \\
    \hline
    && \emph{LP 791-18 d} & \emph{LP 791-18 c} \\
    Inclination, $i$ (deg) & - & $89.34\pm 0.41$$^\textsc{*}$ & $89.94\pm 0.05$ \\
    Semimajor axis, $a$ (au) & - & $0.01992\pm 0.00024$$^\textsc{*}$ & $0.02961^{+0.00035}_{-0.00036}$ \\
    Planet radius, $R_{p}$ (R$_{\oplus}$) & - & $1.032^{+0.044}_{-0.043}$$^\textsc{*}$ & $2.488\pm 0.096$ \\
    Planet mass, $M_{p}$ (M$_{\oplus}$), [Undamped $e$ prior] & - & $0.98\pm 0.30$ & $8.87^{+1.43}_{-1.46}$ \\ 
    Planet mass, $M_{p}$ (M$_{\oplus}$), [Damped $e$ prior] & - & $0.91\pm 0.19$ & $7.16^{+0.65}_{-0.65}$ \\ 
    Bulk density, $\rho$ (g/cm$^3$), [Undamped $e$ prior] & - & $4.92\pm 1.63$ & $3.18\pm 0.64$ \\
    Bulk density, $\rho$ (g/cm$^3$), [Damped $e$ prior] & - & $4.56\pm 1.12$ & $2.56\pm 0.38$ \\
    Eccentricity, $e$, [Undamped $e$ prior] & - & $0.056^{+0.015}_{-0.014}$ & $0.062^{+0.017}_{-0.014}$ \\
    Eccentricity, $e$, [Damped $e$ prior] & - & $0.0011^{+0.0010}_{-0.0008}$ & $0.0001^{+0.0001}_{-0.0001}$ \\
    \hline
    \multicolumn{4}{l}{\footnotesize{$^\textsc{1}$ Centered on the reported period values of P23}} \\
    \multicolumn{4}{l}{\footnotesize{$^\textsc{2}$ Centered on the reported $t_0$ values of P23, adjusted for BJD - 2,458,546}} \\
    \multicolumn{4}{l}{\footnotesize{$^\textsc{*}$ Value from \cite{Peterson2023}}} \\
  \end{tabular}
\end{table*}




\section{Discussion} \label{sec:Discussion}

\subsection{Tidal Heating Constraints} \label{sec:tidal heating}
In the case where LP 791-18 d's free eccentricity is damped and $e_d$ = 0.0015 $\pm$ 0.00014 as reported in P23, then the tidal heat flux at the planet's surface is $10^{-3}$ times the insolation flux. This is still potentially significant for volcanic activity, however. P23 performs detailed modeling of the planet's internal energy budget in this case, including the effects of silicate melting in the planet's interior, and finds that a tidal heat flux per unit mass similar to Io is probable if LP 791-18 d has an Earth-like mantle composition and rheology. While it is possible that LP 791-18 d maintains this forced eccentricity due to dynamical interactions with LP 791-18 c, we find that the TTV observations are not sensitive to eccentricities this small.

\begin{figure}
  \includegraphics[width=8.5cm]{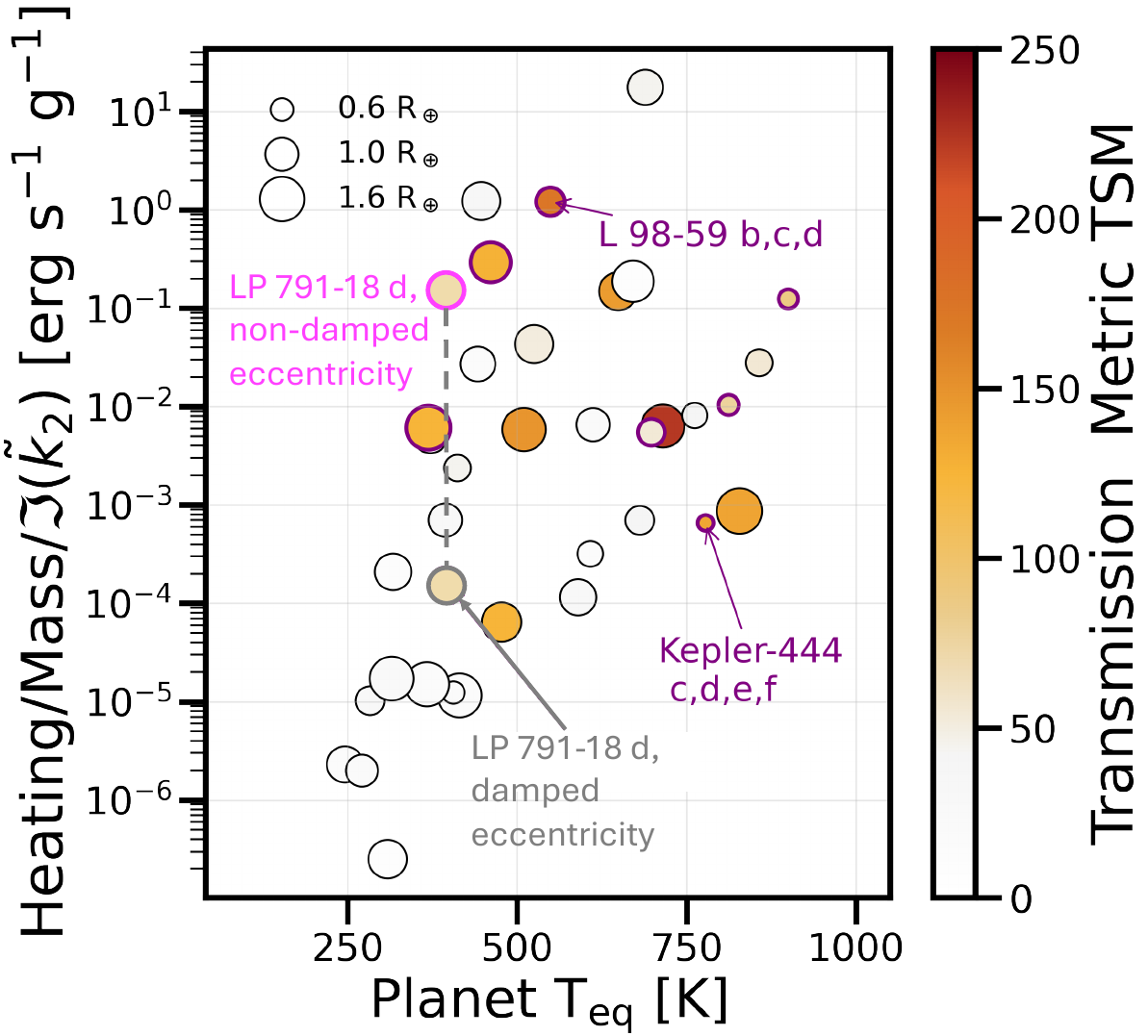}
  \caption{Planetary equilibrium temperature versus tidal heat flux per unit mass, adapted from Figure 5 of \cite{Seligman2023}, showing the most promising rocky planets for tidal volcanism. Heat fluxes are normalized by the uncertain tidal quality parameters, point sizes are scaled by planet size, and color indicates the favorability for atmospheric characterization through transmission spectroscopy \citep{Kempton2018}. The difference in tidal heat flux per unit mass for LP 791-18 d spans more than three orders of magnitude for the damped versus non-damped eccentricity states.}
  \label{fig:tidal_heat_fig}
\end{figure}

If we instead consider the larger eccentricity value for LP 791-18 d from our non-damped TTV retrieval (Table \ref{tab:results}), then the implied tidal heat flux would be much higher than previously reported in P23, with significant implications for the potential outgassing of a secondary atmosphere. 
In this scenario, the tidal heat flux at the surface of LP 791-18 d increases to a value comparable to the insolation flux, and the tidal heat flux per unit mass jumps by more than three orders of magnitude (see Figure \ref{fig:tidal_heat_fig}). Using Equation 1 from \cite{Millholland2020}, we calculate a corresponding $T_\mathrm{int}\approx145$~K for $Q_d$ = $10^4$, compared to a planetary $T_\mathrm{eq}$ of 395 K. In this scenario LP 791-18 d would be comparable to L 98-59 d in terms of tidal heat flux per unit mass \citep{Demangeon2021,Seligman2023}.

\subsection{Impact on Secondary Eclipse Timing}   \label{sec:JWST obs}

\begin{figure}
\begin{center}
  \includegraphics[width=8.5cm]{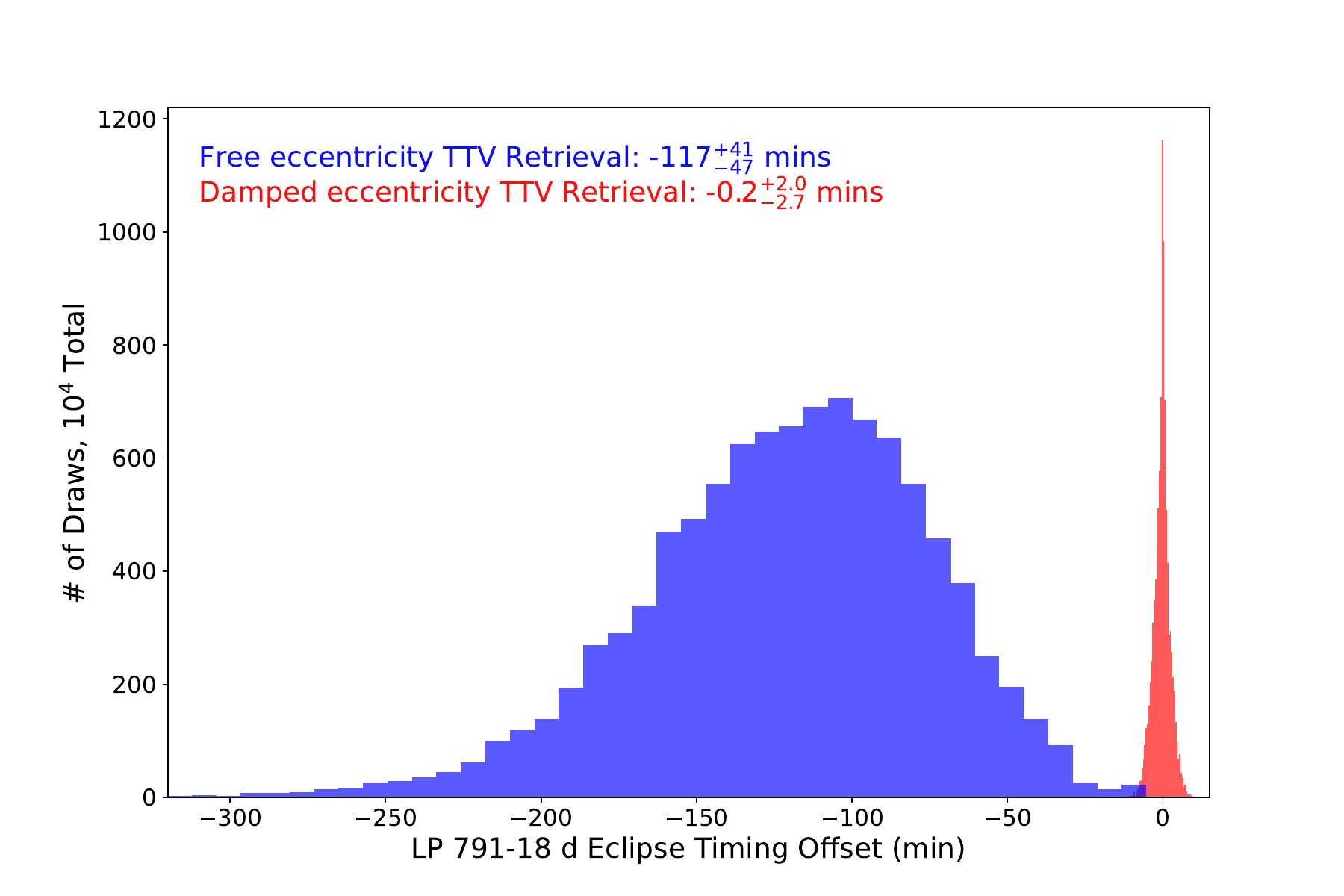}
  \caption{Distributions for the eclipse timing offsets of LP 791-18 d, using $10^4$ draws from the period, e, and $\omega$ distributions of the planets in the damped (red) and non-damped (blue) free eccentricity TTV retrievals.}
  \label{fig:EclipseFig}
\end{center}
\end{figure}

The predicted timing offset of the secondary eclipse relative to the prediction for a planet on a circular orbit is given by $\Delta t=2Pe$cos$\omega/\pi$ \citep{Deming2005}.  This means that if the higher eccentricity for LP 791-18 d preferred by our free retrieval is correct, it could result in a significantly larger timing offset than the damped state prediction.  This has potentially significant implications for upcoming JWST secondary eclipse observations of this planet in program GO 6457, which have a duration of 4.62 hours per observation, with five total eclipse observations currently scheduled. In Figure \ref{fig:EclipseFig}, we compare the distribution of predicted eclipse timing offsets for LP 791-18 d from the tidally damped and free eccentricity TTV retrievals. Notably, with $e_d$ = 0.056 $\pm$ 0.015, the most probable eclipse timing offset is nearly two hours earlier than the value for the damped fits, with a 1$\sigma$ uncertainty window of approximately 45 minutes in either direction. This means that it is possible that the secondary eclipse could occur prior to the start of the JWST observing window.  In contrast to this prediction, our tidally damped fit predicts an eclipse timing offset of -0.2 $^{+2.0}_{-2.7}$ minutes. If this JWST program detects eclipses of planet d with a measured timing offset of more than $10-15$ minutes, it would provide direct observational confirmation that this system is not in the tidally damped equilibrium state. There are no currently scheduled eclipse observations of LP 791-18 c, but the equivalent eclipse timing offset distributions for the non-damped and damped state eccentricities are -273$^{+66}_{-81}$ minutes and 0.0 $^{+0.3}_{-0.4}$ minutes, respectively. 


\subsection{Bulk Density Constraints} \label{sec:bulk densities}

Our new Palomar transit observations of planet c reduce the uncertainty on the planet-to-star radius ratio (R$_p$/ R$_{*}$) by nearly 4$\times$ relative to the value reported in P23. Unfortunately, most of the current uncertainty in the planet radius is driven by the stellar radius uncertainty, so our updated planet radius value (shown in Table \ref{tab:results}) is only minimally shifted relative to P23.  Our mass constraint for the tidally damped case is similar to the value reported by P23, although we prefer a slightly higher mass value for the free eccentricity fit as discussed in \S\ref{sec:non-damped}.  As noted by P23, this planet must have a modest hydrogen-rich envelope; this remains true for either of our updated mass values, although the mass fraction of the envelope is slightly lower when we use the higher mass value from the free eccentricity fit ($\sim$2\%, compared to $\sim$2.5\% atmospheric mass fraction for the lower mass from the damped eccentricity retrieval, \cite{lopez_fortney_2014} Table 2).

\begin{figure}
  \includegraphics[width=8.5cm]{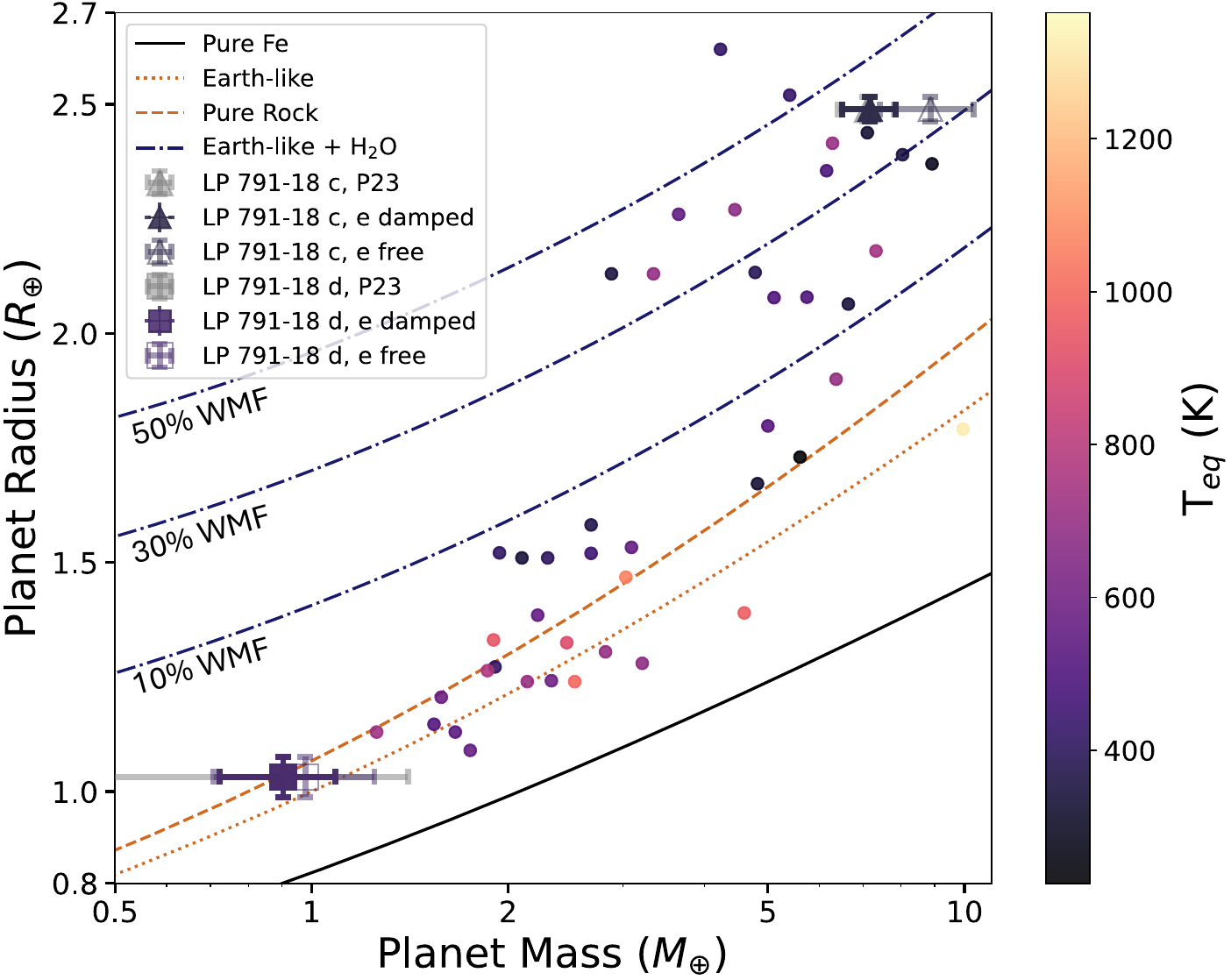}
  \caption{Mass-radius diagram for LP 791-18 d (square) and c (triangle), with updated measurements from this work (filled markers from the damped eccentricity TTV results, open markers for the free eccentricity TTV results) compared to the mass constraints from P23 (gray markers). Filled circles represent all small ($<$ 3 R$_{\oplus}$) planets orbiting M dwarfs (T$_*$ $<$ 3900 K) with masses and radii measured to better than 3$\sigma$, based on the NASA Exoplanet Archive list of confirmed planets as of Nov 2024. The predicted equilibrium temperatures of the planets are indicated by the point color. For comparison, we also plot Earth-like water-rich mass-radius curves (for 10, 30, and 50\% water mass fractions) from \cite{Aguichine2021} and pure iron, Earth-like, and rocky iso-density curves from \cite{zeng_2016}.
  }
  \label{fig:mass_radius}
\end{figure}

Our updated fits also result in a significantly improved mass constraint for LP 791-18 d. P23 reported a mass of 0.9 $^{+0.5}_{-0.4}$ $M_{\oplus}$, while we find a value of $0.91\pm0.19$ $M_{\oplus}$ for the damped case and $0.98^{+0.29}_{-0.25}$ $M_{\oplus}$ for the free eccentricity case. This allows us to better constrain the planetary bulk density (Figure \ref{fig:mass_radius}), which we find is consistent with an Earth-like rock-iron composition within +1$\sigma$ across both fits.  Our best-fit mass for the tidally damped case is slightly lower than the prediction for an Earth-like bulk composition, while the value from our free eccentricity fit is a close match to an Earth-like bulk density.

\section{Summary and Conclusions} \label{sec:Conclusions}

In this study, we present updated constraints on the properties of the temperate Earth-sized  planet LP 791-18 d, and its sub-Neptune sized neighbor LP 791-18 c. We collected and analyzed three new transit observations of LP 791-18 c from Palomar Observatory, which extends the TTV baseline by $\sim$3 years and yields the most precise TTV observations of this system to date (6, 4, and 3~s timing precisions). We use these observations and an updated TTV modeling framework to obtain new constraints on the masses and eccentricities of LP 791-18 c and d, and an updated $R_p$/$R_*$ constraint for LP 791-18~c, which is limited in improving the planet radius constraint by our knowledge of the stellar radius. An updated stellar characterization is beyond the scope of this work, but improved knowledge of the stellar radius for LP 791-18 would result in significant improvements to the radius constraint for LP 791-18 c. We summarize our main conclusions below.

We find that the TTV retrievals with a damped-state eccentricity prior are sensitive to the chosen parameterization for $e$ and $\omega$. When we fit for $e$cos($\omega$) and $e$sin($\omega$) following P23, it results in a small but non-zero eccentricity for both planets due to the effective eccentricity prior bias near zero as described in \cite{exofast} and \cite{Ford2006}. We eliminate this effect by utilizing the $\sqrt{e}$cos($\omega$) and $\sqrt{e}$sin($\omega$) parameterization instead, and find that the TTV observations are consistent with zero orbital eccentricity when a damped state prior is applied. When we repeat our fits without this damped eccentricity prior, the observed TTVs yield moderate non-zero eccentricities for LP 791-18 d and c ($e_d$ = 0.056$^{+0.015}_{-0.014}$ and $e_c$ = 0.062$^{+0.017}_{-0.014}$). For these eccentricities to be maintained on timescales relevant for the age of the system (> 1 Gyr), the tidal quality factors must be larger than is typically assumed for Earth-like planets ($Q_d \geq 10^4$) and sup-Neptunes ($Q_c \geq 10^3$), though not outside the range of predicted values from modeling \citep{Tobie2019} or constraints from population level analysis \citep{Millholland2019}. It is also possible that any non-zero free eccentricity for LP 791-18 c or d might be caused by a more recent dynamical disturbance such as a stellar flyby. 

When using the damped eccentricity prior, we find that the masses of LP 791-18 d and c are consistent with previous constraints reported in P23.  However, the addition of our new TTV observations reduces the uncertainty in planet mass for LP 791-18 d by more than a factor of 2. Our updated mass constraint for LP 791-18 d from this fit is lower than an Earth-like value, but still consistent within 1$\sigma$.  As a result of the known mass-eccentricity degeneracy in TTV models \citep{Lithwick_2012}, our free eccentricity TTV retrieval prefers slightly larger masses for both LP 791-18 d and c, making LP 791-18 d more consistent with an Earth-like bulk composition but not fundamentally changing the interpretation of the planetary compositions from bulk density. 

Our two fits result in significantly different predictions for LP 791-18 d's tidal heat flux. If the planet's eccentricity is close to zero as suggested by the tidally damped fits, this tidal heat flux would be relatively low, though still potentially significant for the planet's composition and evolution as reported in P23.  If the higher eccentricity preferred in the free retrieval is correct, the implied tidal heat flux could be orders of magnitude larger.  We show that upcoming JWST observations of LP 791-18 d's secondary eclipse can easily differentiate between these two scenarios. As long as the eclipse occurs within the JWST observational window, its timing can be used to obtain a significantly tighter constraint on LP 791-18 d's orbital eccentricity, which will in turn result in improved constraints on the masses of both planets in the TTV fit. If the system is in the tidally damped state, JWST observations will be sensitive to much lower ($<$0.01) eccentricities than are detectable in the current TTV data. If JWST instead shows that LP 791-18 d has retained some free eccentricity despite ongoing tidal damping, these observations can be used to constrain its tidal quality factor.

\section{Acknowledgments}

We thank the Palomar Observatory telescope operators, support astronomers, hospitality, and administrative staff, without whom this research would not have been possible. We are especially grateful to Paul Nied, Carolyn Heffner, Kathleen Koviak, Diana Roderick, and Rigel Rafto who supported our observations of LP 791-18 c, and to Monastery keeper Jeff. We thank all of the Palomar Monastery chefs who made observing at Palomar a uniquely pleasant experience, and are sorely missed. Part of this program was supported by JPL Hale telescope time allocations. We are thankful to the PARVI team and Palomar Observatory directorate, especially Chas Beichman, Aurora Kesseli, and Andy Boden for their gracious support of the Palomar TTV survey program during periods requiring quick readjustment of the 200-inch observing schedule. 

This research was supported from the Wilf Family Discovery Fund in Space and Planetary Science established by the Wilf Family Foundation.  This research has made use of the NASA Exoplanet Archive \citep{ps} and the Exoplanet Follow-up Observation Program website, which are operated by the California Institute of Technology, under contract with the National Aeronautics and Space Administration under the Exoplanet Exploration Program. The research made use of the Swarthmore transit finder online tool \citep{SwarthmoreTTF}. 

\software{\texttt{exoplanet} \citep{exoplanet:joss,
exoplanet:zenodo} and its dependencies \citep{exoplanet:foremanmackey17,
exoplanet:foremanmackey18, exoplanet:agol20, exoplanet:arviz,
exoplanet:astropy13, exoplanet:astropy18, exoplanet:luger18, exoplanet:pymc3,
exoplanet:theano}
\texttt{astropy} \citep{astropy},
\texttt{scipy} \citep{scipy},
\texttt{numpy} \citep{numpy},
\texttt{matplotlib} \citep{matplotlib},
\texttt{rebound} \citep{Tamayo2020}, 
\texttt{reboundx} \citep{Lu_2023},
\texttt{BATMAN} \citep{batman},
\texttt{emcee} \citep{emcee},
\texttt{corner} \citep{corner},
\texttt{TTVFast} \citep{TTVFast},
\texttt{lightkurve} \citep{lightkurve}, and
\texttt{Claude 3.5 Sonnet.}}

\facilities{ADS, NASA Exoplanet Archive, TESS, Palomar 200-inch (WIRC)}

%





\appendix

\section{Transit Times and Posterior Probability Distributions} \label{tab:Observed transits}

\begin{table*}[htbp]
\centering
\begin{tabular}{lcccc}
\hline
Planet & Transit \# & Midtime (BJD - 2458540) & $+1\sigma$ & $-1\sigma$ \\
\hline
LP 791-18 d & 0 & 6.3789 & 0.0004 & 0.0004 \\
LP 791-18 d & 1 & 9.1314 & 0.0003 & 0.0003 \\
LP 791-18 d & 2 & 11.8842 & 0.0003 & 0.0003 \\
LP 791-18 d & ... & ... & ... & ... \\
LP 791-18 d & 1433 & 3952.0684 & 0.0099 & 0.0117 \\
LP 791-18 d & 1434 & 3954.8219 & 0.0099 & 0.0117 \\
LP 791-18 d & 1435 & 3957.5772 & 0.0101 & 0.0119 \\
LP 791-18 c & 0 & 6.5092 & 0.0001 & 0.0001 \\
LP 791-18 c & 1 & 11.4992 & 0.0001 & 0.0001 \\
LP 791-18 c & 2 & 16.4892 & 0.0001 & 0.0001 \\
LP 791-18 c & ... & ... & ... & ... \\
LP 791-18 c & 789 & 3943.5488 & 0.0020 & 0.0022 \\
LP 791-18 c & 790 & 3948.5389 & 0.0020 & 0.0023 \\
LP 791-18 c & 791 & 3953.5286 & 0.0020 & 0.0022 \\
\hline
\end{tabular}
\caption{Predicted transit times and uncertainties for LP 791-18 c and d from the TTV fit with a tidally damped eccentricity prior, through January 1, 2030. A subset of rows are depicted here for conciseness. The entirety of this table is provided in the arXiv source code.}
\label{tab:Predicted transits damped}
\end{table*}

\begin{table*}[htbp]
\centering
\begin{tabular}{lcccc}
\hline
Planet & Transit \# & Midtime (BJD - 2458540) & $+1\sigma$ & $-1\sigma$ \\
\hline
LP 791-18 d & 0 & 6.3778 & 0.0007 & 0.0007 \\
LP 791-18 d & 1 & 9.1303 & 0.0006 & 0.0006 \\
LP 791-18 d & 2 & 11.8835 & 0.0005 & 0.0004 \\
LP 791-18 d & ... & ... & ... & ... \\
LP 791-18 d & 1433 & 3952.0897 & 0.0678 & 0.0590 \\
LP 791-18 d & 1434 & 3954.8431 & 0.0677 & 0.0589 \\
LP 791-18 d & 1435 & 3957.5981 & 0.0665 & 0.0581 \\
LP 791-18 c & 0 & 6.5093 & 0.0001 & 0.0001 \\
LP 791-18 c & 1 & 11.4992 & 0.0002 & 0.0002 \\
LP 791-18 c & 2 & 16.4892 & 0.0004 & 0.0004 \\
LP 791-18 c & ... & ... & ... & ... \\
LP 791-18 c & 789 & 3943.5464 & 0.1092 & 0.1125 \\
LP 791-18 c & 790 & 3948.5363 & 0.1094 & 0.1126 \\
LP 791-18 c & 791 & 3953.5259 & 0.1095 & 0.1128 \\
\hline
\end{tabular}
\caption{The same as Table \ref{tab:Predicted transits damped}, but for the non-tidally damped free eccentricity TTV fit.}
\label{tab:Predicted transits undamped}
\end{table*}

\begin{figure*}
\begin{center}
  \includegraphics[width=18.0cm]{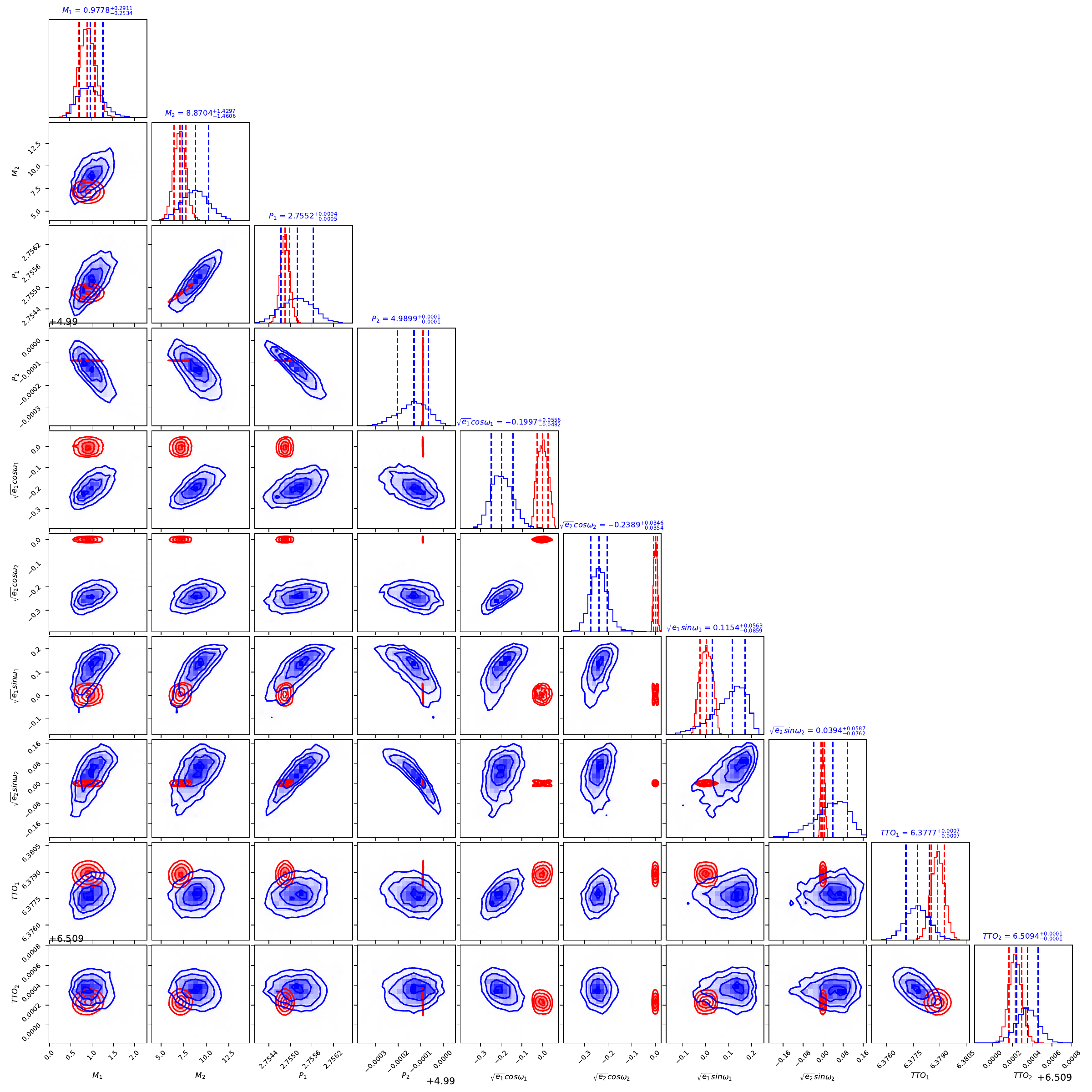}
  \caption{Corner plot of posteriors for TTV model parameters for LP 791-18 d and c from the damped eccentricity (red) and undamped (blue) versions of the TTV fit, made with the \texttt{corner} package \citep{corner}. Fit parameters included planet-to-star mass ratios but we have converted these distributions into units of Earth masses for ease of reference. Columns labels are displayed for the undamped fit results.}
  \label{fig:FullCorner_Comparison}
\end{center}
\end{figure*}


\bibliography{mainLP791}{}

\begin{thebibliography}{}
\expandafter\ifx\csname natexlab\endcsname\relax\def\natexlab#1{#1}\fi
\providecommand{\url}[1]{\href{#1}{#1}}
\providecommand{\dodoi}[1]{doi:~\href{http://doi.org/#1}{\nolinkurl{#1}}}
\providecommand{\doeprint}[1]{\href{http://ascl.net/#1}{\nolinkurl{http://ascl.net/#1}}}
\providecommand{\doarXiv}[1]{\href{https://arxiv.org/abs/#1}{\nolinkurl{https://arxiv.org/abs/#1}}}

\bibitem[{{Agol} {et~al.}(2020){Agol}, {Luger}, \& {Foreman-Mackey}}]{exoplanet:agol20}
{Agol}, E., {Luger}, R., \& {Foreman-Mackey}, D. 2020, \aj, 159, 123, \dodoi{10.3847/1538-3881/ab4fee}

\bibitem[{{Agol} {et~al.}(2021){Agol}, {Dorn}, {Grimm}, {Turbet}, {Ducrot}, {Delrez}, {Gillon}, {Demory}, {Burdanov}, {Barkaoui}, {Benkhaldoun}, {Bolmont}, {Burgasser}, {Carey}, {de Wit}, {Fabrycky}, {Foreman-Mackey}, {Haldemann}, {Hernandez}, {Ingalls}, {Jehin}, {Langford}, {Leconte}, {Lederer}, {Luger}, {Malhotra}, {Meadows}, {Morris}, {Pozuelos}, {Queloz}, {Raymond}, {Selsis}, {Sestovic}, {Triaud}, \& {Van Grootel}}]{Agol_2021}
{Agol}, E., {Dorn}, C., {Grimm}, S.~L., {et~al.} 2021, \psj, 2, 1, \dodoi{10.3847/PSJ/abd022}

\bibitem[{{Aguichine} {et~al.}(2021){Aguichine}, {Mousis}, {Deleuil}, \& {Marcq}}]{Aguichine2021}
{Aguichine}, A., {Mousis}, O., {Deleuil}, M., \& {Marcq}, E. 2021, \apj, 914, 84, \dodoi{10.3847/1538-4357/abfa99}

\bibitem[{{Astropy Collaboration} {et~al.}(2013){Astropy Collaboration}, {Robitaille}, {Tollerud}, {Greenfield}, {Droettboom}, {Bray}, {Aldcroft}, {Davis}, {Ginsburg}, {Price-Whelan}, {Kerzendorf}, {Conley}, {Crighton}, {Barbary}, {Muna}, {Ferguson}, {Grollier}, {Parikh}, {Nair}, {Unther}, {Deil}, {Woillez}, {Conseil}, {Kramer}, {Turner}, {Singer}, {Fox}, {Weaver}, {Zabalza}, {Edwards}, {Azalee Bostroem}, {Burke}, {Casey}, {Crawford}, {Dencheva}, {Ely}, {Jenness}, {Labrie}, {Lim}, {Pierfederici}, {Pontzen}, {Ptak}, {Refsdal}, {Servillat}, \& {Streicher}}]{exoplanet:astropy13}
{Astropy Collaboration}, {Robitaille}, T.~P., {Tollerud}, E.~J., {et~al.} 2013, \aap, 558, A33, \dodoi{10.1051/0004-6361/201322068}

\bibitem[{{Astropy Collaboration} {et~al.}(2018{\natexlab{a}}){Astropy Collaboration}, {Price-Whelan}, {Sip{\H o}cz}, {G{\"u}nther}, {Lim}, {Crawford}, {Conseil}, {Shupe}, {Craig}, {Dencheva}, {Ginsburg}, {VanderPlas}, {Bradley}, {P{\'e}rez-Su{\'a}rez}, {de Val-Borro}, {Aldcroft}, {Cruz}, {Robitaille}, {Tollerud}, {Ardelean}, {Babej}, {Bach}, {Bachetti}, {Bakanov}, {Bamford}, {Barentsen}, {Barmby}, {Baumbach}, {Berry}, {Biscani}, {Boquien}, {Bostroem}, {Bouma}, {Brammer}, {Bray}, {Breytenbach}, {Buddelmeijer}, {Burke}, {Calderone}, {Cano Rodr{\'{\i}}guez}, {Cara}, {Cardoso}, {Cheedella}, {Copin}, {Corrales}, {Crichton}, {D'Avella}, {Deil}, {Depagne}, {Dietrich}, {Donath}, {Droettboom}, {Earl}, {Erben}, {Fabbro}, {Ferreira}, {Finethy}, {Fox}, {Garrison}, {Gibbons}, {Goldstein}, {Gommers}, {Greco}, {Greenfield}, {Groener}, {Grollier}, {Hagen}, {Hirst}, {Homeier}, {Horton}, {Hosseinzadeh}, {Hu}, {Hunkeler}, {Ivezi{\'c}}, {Jain}, {Jenness}, {Kanarek}, {Kendrew}, {Kern}, {Kerzendorf}, {Khvalko}, {King}, {Kirkby},
  {Kulkarni}, {Kumar}, {Lee}, {Lenz}, {Littlefair}, {Ma}, {Macleod}, {Mastropietro}, {McCully}, {Montagnac}, {Morris}, {Mueller}, {Mumford}, {Muna}, {Murphy}, {Nelson}, {Nguyen}, {Ninan}, {N{\"o}the}, {Ogaz}, {Oh}, {Parejko}, {Parley}, {Pascual}, {Patil}, {Patil}, {Plunkett}, {Prochaska}, {Rastogi}, {Reddy Janga}, {Sabater}, {Sakurikar}, {Seifert}, {Sherbert}, {Sherwood-Taylor}, {Shih}, {Sick}, {Silbiger}, {Singanamalla}, {Singer}, {Sladen}, {Sooley}, {Sornarajah}, {Streicher}, {Teuben}, {Thomas}, {Tremblay}, {Turner}, {Terr{\'o}n}, {van Kerkwijk}, {de la Vega}, {Watkins}, {Weaver}, {Whitmore}, {Woillez}, {Zabalza}, \& {Astropy Contributors}}]{exoplanet:astropy18}
{Astropy Collaboration}, {Price-Whelan}, A.~M., {Sip{\H o}cz}, B.~M., {et~al.} 2018{\natexlab{a}}, \aj, 156, 123, \dodoi{10.3847/1538-3881/aabc4f}

\bibitem[{{Astropy Collaboration} {et~al.}(2018{\natexlab{b}}){Astropy Collaboration}, {Price-Whelan}, {Sip{\H{o}}cz}, {G{\"u}nther}, {Lim}, {Crawford}, {Conseil}, {Shupe}, {Craig}, {Dencheva}, {Ginsburg}, {VanderPlas}, {Bradley}, {P{\'e}rez-Su{\'a}rez}, {de Val-Borro}, {Aldcroft}, {Cruz}, {Robitaille}, {Tollerud}, {Ardelean}, {Babej}, {Bach}, {Bachetti}, {Bakanov}, {Bamford}, {Barentsen}, {Barmby}, {Baumbach}, {Berry}, {Biscani}, {Boquien}, {Bostroem}, {Bouma}, {Brammer}, {Bray}, {Breytenbach}, {Buddelmeijer}, {Burke}, {Calderone}, {Cano Rodr{\'\i}guez}, {Cara}, {Cardoso}, {Cheedella}, {Copin}, {Corrales}, {Crichton}, {D'Avella}, {Deil}, {Depagne}, {Dietrich}, {Donath}, {Droettboom}, {Earl}, {Erben}, {Fabbro}, {Ferreira}, {Finethy}, {Fox}, {Garrison}, {Gibbons}, {Goldstein}, {Gommers}, {Greco}, {Greenfield}, {Groener}, {Grollier}, {Hagen}, {Hirst}, {Homeier}, {Horton}, {Hosseinzadeh}, {Hu}, {Hunkeler}, {Ivezi{\'c}}, {Jain}, {Jenness}, {Kanarek}, {Kendrew}, {Kern}, {Kerzendorf}, {Khvalko}, {King}, {Kirkby},
  {Kulkarni}, {Kumar}, {Lee}, {Lenz}, {Littlefair}, {Ma}, {Macleod}, {Mastropietro}, {McCully}, {Montagnac}, {Morris}, {Mueller}, {Mumford}, {Muna}, {Murphy}, {Nelson}, {Nguyen}, {Ninan}, {N{\"o}the}, {Ogaz}, {Oh}, {Parejko}, {Parley}, {Pascual}, {Patil}, {Patil}, {Plunkett}, {Prochaska}, {Rastogi}, {Reddy Janga}, {Sabater}, {Sakurikar}, {Seifert}, {Sherbert}, {Sherwood-Taylor}, {Shih}, {Sick}, {Silbiger}, {Singanamalla}, {Singer}, {Sladen}, {Sooley}, {Sornarajah}, {Streicher}, {Teuben}, {Thomas}, {Tremblay}, {Turner}, {Terr{\'o}n}, {van Kerkwijk}, {de la Vega}, {Watkins}, {Weaver}, {Whitmore}, {Woillez}, {Zabalza}, \& {Astropy Contributors}}]{astropy}
{Astropy Collaboration}, {Price-Whelan}, A.~M., {Sip{\H{o}}cz}, B.~M., {et~al.} 2018{\natexlab{b}}, \aj, 156, 123, \dodoi{10.3847/1538-3881/aabc4f}

\bibitem[{{Banerjee} {et~al.}(2024){Banerjee}, {Barstow}, {Gressier}, {Espinoza}, {Sing}, {Allen}, {Birkmann}, {Challener}, {Crouzet}, {Haswell}, {Lewis}, {Lewis}, \& {Yang}}]{Banerjee2024}
{Banerjee}, A., {Barstow}, J.~K., {Gressier}, A., {et~al.} 2024, \apjl, 975, L11, \dodoi{10.3847/2041-8213/ad73d0}

\bibitem[{{Clausen} \& {Tilgner}(2015)}]{Clausen2015}
{Clausen}, N., \& {Tilgner}, A. 2015, \aap, 584, A60, \dodoi{10.1051/0004-6361/201526082}

\bibitem[{{Crossfield} {et~al.}(2019){Crossfield}, {Waalkes}, {Newton}, {Narita}, {Muirhead}, {Ment}, {Matthews}, {Kraus}, {Kostov}, {Kosiarek}, {Kane}, {Isaacson}, {Halverson}, {Gonzales}, {Everett}, {Dragomir}, {Collins}, {Chontos}, {Berardo}, {Winters}, {Winn}, {Scott}, {Rojas-Ayala}, {Rizzuto}, {Petigura}, {Peterson}, {Mocnik}, {Mikal-Evans}, {Mehrle}, {Matson}, {Kuzuhara}, {Irwin}, {Huber}, {Huang}, {Howell}, {Howard}, {Hirano}, {Fulton}, {Dupuy}, {Dressing}, {Dalba}, {Charbonneau}, {Burt}, {Berta-Thompson}, {Benneke}, {Watanabe}, {Twicken}, {Tamura}, {Schlieder}, {Seager}, {Rose}, {Ricker}, {Quintana}, {L{\'e}pine}, {Latham}, {Kotani}, {Jenkins}, {Hori}, {Colon}, \& {Caldwell}}]{Crossfield_2019}
{Crossfield}, I. J.~M., {Waalkes}, W., {Newton}, E.~R., {et~al.} 2019, \apjl, 883, L16, \dodoi{10.3847/2041-8213/ab3d30}

\bibitem[{{Dai} {et~al.}(2024){Dai}, {Howard}, {Halverson}, {Orell-Miquel}, {Pall{\'e}}, {Isaacson}, {Fulton}, {Price}, {Plotnykov}, {Rogers}, {Valencia}, {Paragas}, {Greklek-McKeon}, {Gomez Barrientos}, {Knutson}, {Petigura}, {Weiss}, {Lee}, {Brinkman}, {Huber}, {Stef{\'a}nsson}, {Masuda}, {Giacalone}, {Lu}, {Kite}, {Hu}, {Gaidos}, {Zhang}, {Rubenzahl}, {Winn}, {Han}, {Beard}, {Holcomb}, {Householder}, {Gilbert}, {Lubin}, {Ong}, {Polanski}, {Saunders}, {Van Zandt}, {Yee}, {Zhang}, {Zink}, {Holden}, {Baker}, {Brodheim}, {Crossfield}, {Deich}, {Edelstein}, {Gibson}, {Hill}, {Jelinsky}, {Kassis}, {Laher}, {Lanclos}, {Lilley}, {Payne}, {Rider}, {Robertson}, {Roy}, {Schwab}, {Shaum}, {Sirk}, {Smith}, {Vandenberg}, {Walawender}, {Wang}, {Wang}, {Wishnow}, {Wright}, {Yeh}, {Caballero}, {Morales}, {Murgas}, {Nagel}, {Reiners}, {Schweitzer}, {Tabernero}, {Zechmeister}, {Spencer}, {Ciardi}, {Clark}, {Lund}, {Caldwell}, {Collins}, {Schwarz}, {Barkaoui}, {Watkins}, {Shporer}, {Narita}, {Fukui}, {Srdoc}, {Latham},
  {Jenkins}, {Ricker}, {Seager}, \& {Vanderspek}}]{Dai2024}
{Dai}, F., {Howard}, A.~W., {Halverson}, S., {et~al.} 2024, \aj, 168, 101, \dodoi{10.3847/1538-3881/ad5a7d}

\bibitem[{Deck {et~al.}(2014)Deck, Agol, Holman, \& Nesvorný}]{TTVFast}
Deck, K.~M., Agol, E., Holman, M.~J., \& Nesvorný, D. 2014, The Astrophysical Journal, 787, 132, \dodoi{10.1088/0004-637x/787/2/132}

\bibitem[{{Demangeon} {et~al.}(2021){Demangeon}, {Zapatero Osorio}, {Alibert}, {Barros}, {Adibekyan}, {Tabernero}, {Antoniadis-Karnavas}, {Camacho}, {Su{\'a}rez Mascare{\~n}o}, {Oshagh}, {Micela}, {Sousa}, {Lovis}, {Pepe}, {Rebolo}, {Cristiani}, {Santos}, {Allart}, {Allende Prieto}, {Bossini}, {Bouchy}, {Cabral}, {Damasso}, {Di Marcantonio}, {D'Odorico}, {Ehrenreich}, {Faria}, {Figueira}, {G{\'e}nova Santos}, {Haldemann}, {Hara}, {Gonz{\'a}lez Hern{\'a}ndez}, {Lavie}, {Lillo-Box}, {Lo Curto}, {Martins}, {M{\'e}gevand}, {Mehner}, {Molaro}, {Nunes}, {Pall{\'e}}, {Pasquini}, {Poretti}, {Sozzetti}, \& {Udry}}]{Demangeon2021}
{Demangeon}, O.~D.~S., {Zapatero Osorio}, M.~R., {Alibert}, Y., {et~al.} 2021, \aap, 653, A41, \dodoi{10.1051/0004-6361/202140728}

\bibitem[{{Deming} {et~al.}(2005){Deming}, {Seager}, {Richardson}, \& {Harrington}}]{Deming2005}
{Deming}, D., {Seager}, S., {Richardson}, L.~J., \& {Harrington}, J. 2005, \nat, 434, 740, \dodoi{10.1038/nature03507}

\bibitem[{Eastman {et~al.}(2013)Eastman, Gaudi, \& Agol}]{exofast}
Eastman, J., Gaudi, B.~S., \& Agol, E. 2013, Publications of the Astronomical Society of the Pacific, 125, 83–112, \dodoi{10.1086/669497}

\bibitem[{{Fischer} {et~al.}(2016){Fischer}, {Anglada-Escude}, {Arriagada}, {Baluev}, {Bean}, {Bouchy}, {Buchhave}, {Carroll}, {Chakraborty}, {Crepp}, {Dawson}, {Diddams}, {Dumusque}, {Eastman}, {Endl}, {Figueira}, {Ford}, {Foreman-Mackey}, {Fournier}, {F{\H{u}}r{\'e}sz}, {Gaudi}, {Gregory}, {Grundahl}, {Hatzes}, {H{\'e}brard}, {Herrero}, {Hogg}, {Howard}, {Johnson}, {Jorden}, {Jurgenson}, {Latham}, {Laughlin}, {Loredo}, {Lovis}, {Mahadevan}, {McCracken}, {Pepe}, {Perez}, {Phillips}, {Plavchan}, {Prato}, {Quirrenbach}, {Reiners}, {Robertson}, {Santos}, {Sawyer}, {Segransan}, {Sozzetti}, {Steinmetz}, {Szentgyorgyi}, {Udry}, {Valenti}, {Wang}, {Wittenmyer}, \& {Wright}}]{Fischer2016}
{Fischer}, D.~A., {Anglada-Escude}, G., {Arriagada}, P., {et~al.} 2016, \pasp, 128, 066001, \dodoi{10.1088/1538-3873/128/964/066001}

\bibitem[{{Ford}(2006)}]{Ford2006}
{Ford}, E.~B. 2006, \apj, 642, 505, \dodoi{10.1086/500802}

\bibitem[{Foreman-Mackey(2016)}]{corner}
Foreman-Mackey, D. 2016, The Journal of Open Source Software, 24, \dodoi{10.21105/joss.00024}

\bibitem[{{Foreman-Mackey}(2018)}]{exoplanet:foremanmackey18}
{Foreman-Mackey}, D. 2018, Research Notes of the American Astronomical Society, 2, 31, \dodoi{10.3847/2515-5172/aaaf6c}

\bibitem[{{Foreman-Mackey} {et~al.}(2017){Foreman-Mackey}, {Agol}, {Ambikasaran}, \& {Angus}}]{exoplanet:foremanmackey17}
{Foreman-Mackey}, D., {Agol}, E., {Ambikasaran}, S., \& {Angus}, R. 2017, \aj, 154, 220, \dodoi{10.3847/1538-3881/aa9332}

\bibitem[{Foreman-Mackey {et~al.}(2013)Foreman-Mackey, Hogg, Lang, \& Goodman}]{emcee}
Foreman-Mackey, D., Hogg, D.~W., Lang, D., \& Goodman, J. 2013, Publications of the Astronomical Society of the Pacific, 125, 306–312, \dodoi{10.1086/670067}

\bibitem[{{Foreman-Mackey} {et~al.}(2021){Foreman-Mackey}, {Luger}, {Agol}, {Barclay}, {Bouma}, {Brandt}, {Czekala}, {David}, {Dong}, {Gilbert}, {Gordon}, {Hedges}, {Hey}, {Morris}, {Price-Whelan}, \& {Savel}}]{exoplanet:joss}
{Foreman-Mackey}, D., {Luger}, R., {Agol}, E., {et~al.} 2021, arXiv e-prints, arXiv:2105.01994.
\newblock \doarXiv{2105.01994}

\bibitem[{Foreman-Mackey {et~al.}(2021)Foreman-Mackey, Savel, Luger, Agol, Czekala, Price-Whelan, Hedges, Gilbert, Bouma, Brandt, \& Barclay}]{exoplanet:zenodo}
Foreman-Mackey, D., Savel, A., Luger, R., {et~al.} 2021, exoplanet-dev/exoplanet v0.5.1, \dodoi{10.5281/zenodo.1998447}

\bibitem[{{Goldreich} \& {Soter}(1966)}]{Goldreich1996}
{Goldreich}, P., \& {Soter}, S. 1966, \icarus, 5, 375, \dodoi{10.1016/0019-1035(66)90051-0}

\bibitem[{{Greklek-McKeon} {et~al.}(2023){Greklek-McKeon}, {Knutson}, {Vissapragada}, {Jontof-Hutter}, {Chachan}, {Thorngren}, \& {Vasisht}}]{GreklekMcKeon2023}
{Greklek-McKeon}, M., {Knutson}, H.~A., {Vissapragada}, S., {et~al.} 2023, \aj, 165, 48, \dodoi{10.3847/1538-3881/ac8553}

\bibitem[{{Gressier} {et~al.}(2024){Gressier}, {Espinoza}, {Allen}, {Sing}, {Banerjee}, {Barstow}, {Valenti}, {Lewis}, {Birkmann}, {Challener}, {Manjavacas}, {Alves de Oliveira}, {Crouzet}, \& {Beck}}]{Gressier2024}
{Gressier}, A., {Espinoza}, N., {Allen}, N.~H., {et~al.} 2024, \apjl, 975, L10, \dodoi{10.3847/2041-8213/ad73d1}

\bibitem[{Harris {et~al.}(2020)Harris, Millman, van~der Walt, Gommers, Virtanen, Cournapeau, Wieser, Taylor, Berg, Smith, Kern, Picus, Hoyer, van Kerkwijk, Brett, Haldane, Fernández~del Río, Wiebe, Peterson, Gérard-Marchant, Sheppard, Reddy, Weckesser, Abbasi, Gohlke, \& Oliphant}]{numpy}
Harris, C.~R., Millman, K.~J., van~der Walt, S.~J., {et~al.} 2020, Nature, 585, 357–362, \dodoi{10.1038/s41586-020-2649-2}

\bibitem[{Hunter(2007)}]{matplotlib}
Hunter, J.~D. 2007, Computing In Science \& Engineering, 9, 90

\bibitem[{{Jackson} {et~al.}(2008){Jackson}, {Barnes}, \& {Greenberg}}]{Jackson2008}
{Jackson}, B., {Barnes}, R., \& {Greenberg}, R. 2008, \mnras, 391, 237, \dodoi{10.1111/j.1365-2966.2008.13868.x}

\bibitem[{{Jensen}(2013)}]{SwarthmoreTTF}
{Jensen}, E. 2013, {Tapir: A web interface for transit/eclipse observability}, Astrophysics Source Code Library, record ascl:1306.007

\bibitem[{{Kempton} {et~al.}(2018){Kempton}, {Bean}, {Louie}, {Deming}, {Koll}, {Mansfield}, {Christiansen}, {L{\'o}pez-Morales}, {Swain}, {Zellem}, {Ballard}, {Barclay}, {Barstow}, {Batalha}, {Beatty}, {Berta-Thompson}, {Birkby}, {Buchhave}, {Charbonneau}, {Cowan}, {Crossfield}, {de Val-Borro}, {Doyon}, {Dragomir}, {Gaidos}, {Heng}, {Hu}, {Kane}, {Kreidberg}, {Mallonn}, {Morley}, {Narita}, {Nascimbeni}, {Pall{\'e}}, {Quintana}, {Rauscher}, {Seager}, {Shkolnik}, {Sing}, {Sozzetti}, {Stassun}, {Valenti}, \& {von Essen}}]{Kempton2018}
{Kempton}, E. M.~R., {Bean}, J.~L., {Louie}, D.~R., {et~al.} 2018, \pasp, 130, 114401, \dodoi{10.1088/1538-3873/aadf6f}

\bibitem[{Kreidberg(2015)}]{batman}
Kreidberg, L. 2015, Publications of the Astronomical Society of the Pacific, 127, 1161

\bibitem[{Kumar {et~al.}(2019)Kumar, Carroll, Hartikainen, \& Martin}]{exoplanet:arviz}
Kumar, R., Carroll, C., Hartikainen, A., \& Martin, O.~A. 2019, The Journal of Open Source Software, \dodoi{10.21105/joss.01143}

\bibitem[{{Lightkurve Collaboration} {et~al.}(2018){Lightkurve Collaboration}, {Cardoso}, {Hedges}, {Gully-Santiago}, {Saunders}, {Cody}, {Barclay}, {Hall}, {Sagear}, {Turtelboom}, {Zhang}, {Tzanidakis}, {Mighell}, {Coughlin}, {Bell}, {Berta-Thompson}, {Williams}, {Dotson}, \& {Barentsen}}]{lightkurve}
{Lightkurve Collaboration}, {Cardoso}, J.~V.~d.~M., {Hedges}, C., {et~al.} 2018, {Lightkurve: Kepler and TESS time series analysis in Python}, Astrophysics Source Code Library.
\newblock \doeprint{1812.013}

\bibitem[{Lithwick {et~al.}(2012)Lithwick, Xie, \& Wu}]{Lithwick_2012}
Lithwick, Y., Xie, J., \& Wu, Y. 2012, The Astrophysical Journal, 761, 122, \dodoi{10.1088/0004-637x/761/2/122}

\bibitem[{{Lopez} \& {Fortney}(2014)}]{lopez_fortney_2014}
{Lopez}, E.~D., \& {Fortney}, J.~J. 2014, \apj, 792, 1, \dodoi{10.1088/0004-637X/792/1/1}

\bibitem[{{Louden} {et~al.}(2023){Louden}, {Laughlin}, \& {Millholland}}]{Louden2023}
{Louden}, E.~M., {Laughlin}, G.~P., \& {Millholland}, S.~C. 2023, \apjl, 958, L21, \dodoi{10.3847/2041-8213/ad0843}

\bibitem[{Lu {et~al.}(2023)Lu, Rein, Tamayo, Hadden, Mardling, Millholland, \& Laughlin}]{Lu_2023}
Lu, T., Rein, H., Tamayo, D., {et~al.} 2023, The Astrophysical Journal, 948, 41, \dodoi{10.3847/1538-4357/acc06d}

\bibitem[{{Luger} {et~al.}(2019){Luger}, {Agol}, {Foreman-Mackey}, {Fleming}, {Lustig-Yaeger}, \& {Deitrick}}]{exoplanet:luger18}
{Luger}, R., {Agol}, E., {Foreman-Mackey}, D., {et~al.} 2019, \aj, 157, 64, \dodoi{10.3847/1538-3881/aae8e5}

\bibitem[{{Millholland}(2019)}]{Millholland2019}
{Millholland}, S. 2019, \apj, 886, 72, \dodoi{10.3847/1538-4357/ab4c3f}

\bibitem[{{Millholland} {et~al.}(2020){Millholland}, {Petigura}, \& {Batygin}}]{Millholland2020}
{Millholland}, S., {Petigura}, E., \& {Batygin}, K. 2020, \apj, 897, 7, \dodoi{10.3847/1538-4357/ab959c}

\bibitem[{{Morley} {et~al.}(2017){Morley}, {Knutson}, {Line}, {Fortney}, {Thorngren}, {Marley}, {Teal}, \& {Lupu}}]{Morley2017}
{Morley}, C.~V., {Knutson}, H., {Line}, M., {et~al.} 2017, \aj, 153, 86, \dodoi{10.3847/1538-3881/153/2/86}

\bibitem[{{NASA Exoplanet Archive}(2024{\natexlab{a}})}]{NASA_exoplanet_archive}
{NASA Exoplanet Archive}. 2024{\natexlab{a}}, Planetary Systems Composite Parameters, Version: 2024-11-26,  NExScI-Caltech/IPAC, \dodoi{10.26133/NEA13}

\bibitem[{{NASA Exoplanet Archive}(2024{\natexlab{b}})}]{ps}
---. 2024{\natexlab{b}}, Planetary Systems, Version: 2024-11-01 12:00,  NExScI-Caltech/IPAC, \dodoi{10.26133/NEA12}

\bibitem[{{Nettelmann} {et~al.}(2011){Nettelmann}, {Fortney}, {Kramm}, \& {Redmer}}]{Nettelmann2011}
{Nettelmann}, N., {Fortney}, J.~J., {Kramm}, U., \& {Redmer}, R. 2011, \apj, 733, 2, \dodoi{10.1088/0004-637X/733/1/2}

\bibitem[{{P{\'e}rez-Gonz{\'a}lez} {et~al.}(2024){P{\'e}rez-Gonz{\'a}lez}, {Greklek-McKeon}, {Vissapragada}, {Saidel}, {Knutson}, {Linssen}, \& {Oklop{\v{c}}i{\'c}}}]{Jorge2024}
{P{\'e}rez-Gonz{\'a}lez}, J., {Greklek-McKeon}, M., {Vissapragada}, S., {et~al.} 2024, \aj, 167, 214, \dodoi{10.3847/1538-3881/ad34b6}

\bibitem[{{Peterson} {et~al.}(2023){Peterson}, {Benneke}, {Collins}, {Piaulet}, {Crossfield}, {Ali-Dib}, {Christiansen}, {Gagn{\'e}}, {Faherty}, {Kite}, {Dressing}, {Charbonneau}, {Murgas}, {Cointepas}, {Almenara}, {Bonfils}, {Kane}, {Werner}, {Gorjian}, {Roy}, {Shporer}, {Pozuelos}, {Socia}, {Cloutier}, {Dietrich}, {Irwin}, {Weiss}, {Waalkes}, {Berta-Thomson}, {Evans}, {Apai}, {Parviainen}, {Pall{\'e}}, {Narita}, {Howard}, {Dragomir}, {Barkaoui}, {Gillon}, {Jehin}, {Ducrot}, {Benkhaldoun}, {Fukui}, {Mori}, {Nishiumi}, {Kawauchi}, {Ricker}, {Latham}, {Winn}, {Seager}, {Isaacson}, {Bixel}, {Gibbs}, {Jenkins}, {Smith}, {Chavez}, {Rackham}, {Henning}, {Gabor}, {Chen}, {Espinoza}, {Jensen}, {Collins}, {Schwarz}, {Conti}, {Wang}, {Kielkopf}, {Mao}, {Horne}, {Sefako}, {Quinn}, {Moldovan}, {Fausnaugh}, {F{\.z}{\.z}r{\'e}sz}, \& {Barclay}}]{Peterson2023}
{Peterson}, M.~S., {Benneke}, B., {Collins}, K., {et~al.} 2023, \nat, 617, 701, \dodoi{10.1038/s41586-023-05934-8}

\bibitem[{{Piaulet-Ghorayeb} {et~al.}(2024){Piaulet-Ghorayeb}, {Benneke}, {Radica}, {Raul}, {Coulombe}, {Ahrer}, {Kubyshkina}, {Howard}, {Krissansen-Totton}, {MacDonald}, {Roy}, {Louca}, {Christie}, {Fournier-Tondreau}, {Allart}, {Miguel}, {Schlichting}, {Welbanks}, {Cadieux}, {Dorn}, {Evans-Soma}, {Fortney}, {Pierrehumbert}, {Lafreni{\`e}re}, {Acu{\~n}a}, {Komacek}, {Innes}, {Beatty}, {Cloutier}, {Doyon}, {Gagnebin}, {Gapp}, \& {Knutson}}]{Piaulet2024}
{Piaulet-Ghorayeb}, C., {Benneke}, B., {Radica}, M., {et~al.} 2024, \apjl, 974, L10, \dodoi{10.3847/2041-8213/ad6f00}

\bibitem[{{Puranam} \& {Batygin}(2018)}]{Puranam2018}
{Puranam}, A., \& {Batygin}, K. 2018, \aj, 155, 157, \dodoi{10.3847/1538-3881/aab09f}

\bibitem[{{Quick} {et~al.}(2020){Quick}, {Roberge}, {Mlinar}, \& {Hedman}}]{Quick2020}
{Quick}, L.~C., {Roberge}, A., {Mlinar}, A.~B., \& {Hedman}, M.~M. 2020, \pasp, 132, 084402, \dodoi{10.1088/1538-3873/ab9504}

\bibitem[{{Raftery}(1995)}]{raftery1995}
{Raftery}, A.~E. 1995, JSTOR, 25, 111–63, \dodoi{10.2307/271063}

\bibitem[{Rein \& Tamayo(2015)}]{REBOUND}
Rein, H., \& Tamayo, D. 2015, Monthly Notices of the Royal Astronomical Society, 452, 376–388, \dodoi{10.1093/mnras/stv1257}

\bibitem[{Salvatier {et~al.}(2016)Salvatier, Wiecki, \& Fonnesbeck}]{exoplanet:pymc3}
Salvatier, J., Wiecki, T.~V., \& Fonnesbeck, C. 2016, PeerJ Computer Science, 2, e55

\bibitem[{{Schwarz}(1978)}]{Schwarz1978}
{Schwarz}, G. 1978, Annals of Statistics, 6, 461

\bibitem[{SciPy(2001)}]{scipy}
SciPy. 2001, {SciPy}: Open source scientific tools for Python.
\newblock \url{http://www.scipy.org/}

\bibitem[{Seligman {et~al.}(2023)Seligman, Feinstein, Lai, Welbanks, Taylor, Becker, Adams, Morgan, \& Bergner}]{Seligman2023}
Seligman, D.~Z., Feinstein, A.~D., Lai, D., {et~al.} 2023, Potential Melting of Extrasolar Planets by Tidal Dissipation.
\newblock \doarXiv{2311.01187}

\bibitem[{{Stefansson} {et~al.}(2017){Stefansson}, {Mahadevan}, {Hebb}, {Wisniewski}, {Huehnerhoff}, {Morris}, {Halverson}, {Zhao}, {Wright}, {O'rourke}, {Knutson}, {Hawley}, {Kanodia}, {Li}, {Hagen}, {Liu}, {Beatty}, {Bender}, {Robertson}, {Dembicky}, {Gray}, {Ketzeback}, {McMillan}, \& {Rudyk}}]{Stefansson2017}
{Stefansson}, G., {Mahadevan}, S., {Hebb}, L., {et~al.} 2017, \apj, 848, 9, \dodoi{10.3847/1538-4357/aa88aa}

\bibitem[{{Tamayo} {et~al.}(2020){Tamayo}, {Rein}, {Shi}, \& {Hernandez}}]{Tamayo2020}
{Tamayo}, D., {Rein}, H., {Shi}, P., \& {Hernandez}, D.~M. 2020, \mnras, 491, 2885, \dodoi{10.1093/mnras/stz2870}

\bibitem[{{Theano Development Team}(2016)}]{exoplanet:theano}
{Theano Development Team}. 2016, arXiv e-prints, abs/1605.02688.
\newblock \url{http://arxiv.org/abs/1605.02688}

\bibitem[{{Tobie} {et~al.}(2019){Tobie}, {Grasset}, {Dumoulin}, \& {Mocquet}}]{Tobie2019}
{Tobie}, G., {Grasset}, O., {Dumoulin}, C., \& {Mocquet}, A. 2019, \aap, 630, A70, \dodoi{10.1051/0004-6361/201935297}

\bibitem[{Vissapragada {et~al.}(2020)Vissapragada, Jontof-Hutter, Shporer, Knutson, Liu, Thorngren, Lee, Chachan, Mawet, Millar-Blanchaer, \& et~al.}]{Vissapragada2020}
Vissapragada, S., Jontof-Hutter, D., Shporer, A., {et~al.} 2020, The Astronomical Journal, 159, 108, \dodoi{10.3847/1538-3881/ab65c8}

\bibitem[{{Wilson} {et~al.}(2003){Wilson}, {Eikenberry}, {Henderson}, {Hayward}, {Carson}, {Pirger}, {Barry}, {Brandl}, {Houck}, {Fitzgerald}, \& {Stolberg}}]{Wilson2003}
{Wilson}, J.~C., {Eikenberry}, S.~S., {Henderson}, C.~P., {et~al.} 2003, in Society of Photo-Optical Instrumentation Engineers (SPIE) Conference Series, Vol. 4841, Instrument Design and Performance for Optical/Infrared Ground-based Telescopes, ed. M.~{Iye} \& A.~F.~M. {Moorwood}, 451--458, \dodoi{10.1117/12.460336}

\bibitem[{{Wright}(2018)}]{Wright2018}
{Wright}, J.~T. 2018, in Handbook of Exoplanets, ed. H.~J. {Deeg} \& J.~A. {Belmonte}, 4, \dodoi{10.1007/978-3-319-55333-7_4}

\bibitem[{Zeng {et~al.}(2016)Zeng, Sasselov, \& Jacobsen}]{zeng_2016}
Zeng, L., Sasselov, D.~D., \& Jacobsen, S.~B. 2016, The Astrophysical Journal, 819, 127, \dodoi{10.3847/0004-637x/819/2/127}

\end{thebibliography}
\bibliographystyle{aasjournal}



\end{document}